\documentclass[useAMS,usenatbib]{mn2e}
\usepackage{deluxetable, color}
\usepackage{amssymb,amsmath}
\usepackage{graphicx}
\usepackage{longtable, float, subfigure}
\usepackage[T1]{fontenc}
\usepackage[latin1]{inputenc}
\def\kms{km s$^{-1}$}
\def\et{{et~al.}}
\def\ha{H$\alpha$}
\def\arcdeg{\ifmmode^\circ\;\else$^\circ$\fi}
\def\arcmin{\ifmmode^\prime\;\else$^\prime$\fi}
\def\arcsec{\ifmmode^{\prime\prime}\;\else$^{\prime\prime}$\fi}
\def\deg{\ifmmode^\circ\;\else$^\circ$\fi}

\def\hi{H\;\!{\sc i}}
\def\vmax{$V_{\rm max}$}

\def\msun{M$_{\sun}$}

\title{Blasting away a dwarf galaxy:  The ``tail'' of ESO 324-G024}
\author[Megan C. Johnson]{Megan C.\ Johnson$^{1}$\thanks{E-mail: megan.johnson@csiro.au}, 
Peter Kamphuis$^{1}$, 
B\"arbel S.\ Koribalski$^{1}$, 
Jing Wang$^{1}$, 
\newauthor
Se-Heon Oh$^{2,3}$, 
Alex Hill$^{1,4}$, 
Shane O'Sullivan$^{5}$, 
Sebastian Haan$^{1}$ 
and Paolo Serra$^{1}$
\\
$^{1}$CSIRO Astronomy and Space Science, Australia Telescope National Facility (ATNF), Epping, NSW\\
$^{2}$The International Centre for Radio Astronomy Research (ICRAR), The University of Western Australia, 35 Stirling Highway, \\
Crawley, Western Australia 6009, Australia\\
$^{3}$ARC Centre of Excellence for All-sky Astrophysics (CAASTRO)\\
$^{4}$Current address: Departments of Physics and Astronomy, Haverford College, Haverford PA 19041 USA\\
$^{5}$Instituto de Astronom\'ia, Universidad Nacional Aut\'onoma de M\'exico (UNAM), A.P.~70-264, 04510 M\'exico, D.F., M\'exico
}

\begin{document}

\maketitle

\begin{abstract}
We 
 present Australia Telescope Compact Array radio data of the dwarf
 irregular galaxy ESO 324-G024 which is seen in projection against the giant, 
 northern lobe of the radio galaxy Centaurus A (Cen A, NGC~5128). The distorted 
 morphology and kinematics of ESO 324-G024, as observed in the 21 cm spectral 
 line emission of neutral hydrogen, indicate disruptions by external 
 forces. We investigate whether tidal interactions and/or ram pressure stripping are responsible 
 for the formation of the \hi\ tail stretching to the northeast of ESO 324-G024 with the latter being most probable. 
 Furthermore, we closely analyze the sub-structure
 of Cen A's polarized radio lobes to ascertain whether ESO 324-G024 is located
 in front, within or behind the northern lobe. Our multi-wavelength, multi-component approach allows us to determine that ESO 324-G024 is most likely behind the northern radio lobe of Cen A.  This result helps to constrain the orientation of the lobe, which is likely inclined to our line of sight by approximately 60$\arcdeg$ if NGC 5128 and ESO 324-G024 are at the same distance.
\end{abstract}

\section{Introduction}\label{sec:intro}

According to concordance simulations employing cold dark matter particle physics with a flat Euclidean geometry, the well-known $\Lambda$CDM cosmology, the massive galaxies we observe today were assembled through the coalescence of many smaller systems.  It was thought up until recently, that the dwarf galaxies observed in the local Universe were the left-over remnants of this earlier epoch of massive galaxy formation.  Hence, dwarf galaxies, which are the most abundant types of galaxies, were 
generally regarded as the `building blocks' of large galaxies 
\citep{hod03, van04}.
   
  There are examples locally of dwarf galaxies being accreted by larger galaxies as in the cases of 
  the Fornax dwarf spheroidal galaxy \citep{yoz12} and the Sagittarius stream \citep{lyn95} who are being accreted by the Milky Way. 
   However, recent studies have shown that the dwarf galaxies we observe in the nearby Universe, may have actually formed later, after the most massive galaxies coalesced \citep{tos03, guo09, pau10, guo11}.  Nevertheless, the properties of dwarf galaxies, in particular their
  low metallicities, low shear, and shallow gravitational potential wells, resemble the conditions in the early Universe when massive galaxies formed and therefore, make excellent laboratories for studying how stars formed in the early Universe.
%

\begin{figure*}
\centering
\includegraphics[scale=.4]{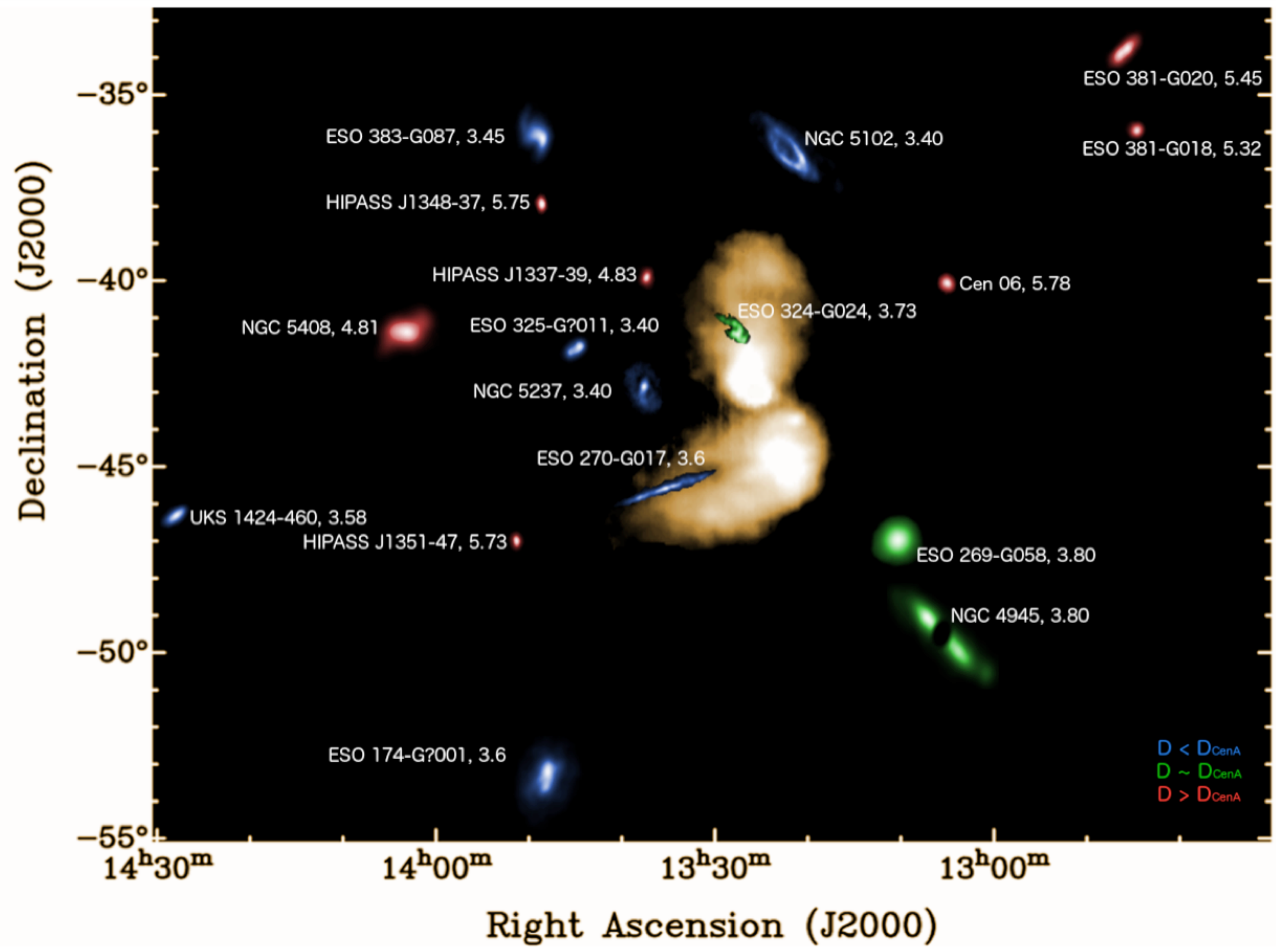}
\caption{The Cen\,A galaxy group: In this composite image the giant radio lobes 
 of the Cen\,A galaxy are shown using the CHIPASS 1.4 GHz (20 cm) continuum emission 
 (Calabretta et al. 2014), while the neighboring gas-rich galaxies are 
 highlighted using the ATCA \hi\ intensity maps from the LVHIS project 
 (Koribalski 2008). The latter are enlarged by a factor ten and color-coded 
 to reflect their relative distances to Cen\,A. The galaxy name and distance
 (in Mpc) is annotated in the figure. --- This composition was inspired by 
 the Chung et al. (2009) image of the gas-rich Virgo cluster galaxies.}\label{fig:1a}
\end{figure*}

An 
example of a large galaxy in the process of accreting its satellite galaxy neighbors is the giant elliptical galaxy NGC 5128, which contains the bright radio source known as Centaurus A (Cen A).  Its enormous radio lobes
 span about 9$\arcdeg \times\ 4\arcdeg$ on the sky \citep{coo65, jun93, mor99, fea11}, which corresponds to a total north--south 
 extent of at least 600 kpc at our adopted distance of 3.8 $\pm$ 0.1 Mpc \citep{har10}. Table \ref{tab:1} lists some global properties for NGC 5128 and ESO 324-G024 and Figure \ref{fig:1a} 
 shows 1.4 GHz 
 radio continuum emission of Cen\,A as mapped with the 64-m Parkes Telescope
 \citep{cal14} as well as the \hi\  emission of the surrounding companion
 galaxies obtained with the Australia Telescope Compact Array (ATCA) as part 
 of the `Local Volume \hi\ Survey' \citep[LVHIS,][]{kor08}.

 NGC 5128 contains
 many features \color{black}including a supermassive black hole in its central core \citep[see e.g.,][and references therein]{cap09}, which is actively accreting material and producing a jet that can be seen in X-ray emission \citep{kra02} and radio continuum \citep{fei82}.  The X-ray jet is highly collimated and extends a projected $\sim$4 kpc (at our adopted distance) northeast from the nucleus of NGC 5128 at an average sky position angle of $\sim$55$\arcdeg$ \citep{sch79, fei82, kra02}.  NGC 5128 has 
 a dust lane that extends across its stellar disk \citep[e.g.][]{mal83} and arcs of stellar shells that stretch nearly 15$\arcmin$ in radii and provide evidence of its merger history \citep[][and references therein]{mal83, mal97, pen02}.  It contains \hi\ \citep{sch94, str10} shells that harbor molecular gas as traced by CO \citep{cha00} and these features extend $\sim$15 kpc, as far as the outermost stellar shells, along the direction of the jet. NGC 5128 is sometimes classified as an S0 object in the Hubble galaxy classification scheme because of its prominent dust lane \citep{dev91}. It is believed in the literature that NGC 5128 is likely in the process of transitioning from a gas-rich spiral galaxy to a relatively gas-poor elliptical galaxy \citep{mal97} or vice versa \citep{hau08}.  

For a comprehensive review of Cen A and its host galaxy NGC 5128 see, e.g., \citet{isr98, fea11, osu13, eil14}.

 From Figure \ref{fig:1a}, we see that the Centaurus A galaxy group is rich in dwarf irregular (dIrr) galaxies. Two objects
 stand out as they appears in projection against the northern radio lobe of Cen A.
 The edge-on gas-rich disk galaxy ESO 270-G017 has a distance that is not well determined as it was derived from the Tully-Fisher relation.  Therefore, it is likely that this galaxy is not even part of the Centaurus A galaxy group as it has a systemic velocity that differs $\sim$300 km s$^{-1}$ from that of NGC 5128.
  
  ESO 324-G024, on the other hand, has a 
distance of 3.73 $\pm$ 0.43 Mpc, as determined from the the tip of the red giant branch (TRGB) \citep{kar02}, which implies that its distance is nearly the same as NGC 5128; ESO 324-G024 is the focus of this paper.
 It lies a projected 104 kpc north of NGC 5128's core, which means ESO 324-G024 could possibly be \emph {inside} the northern radio lobe.  We explore the implications for this unique scenario and aim to understand
 whether tidal interactions and/or ram pressure stripping are responsible for its peculiar \hi\ morphology.  
 
Several other dwarf galaxies reside close to Cen\,A, although ESO 324-G024 is likely the closest dIrr to NGC 5128. The closest dwarf spheroidal (dSph) galaxies to NGC 5128 are possibly KKs\,55 and 
 KK\,197, separated by 0.66$\arcdeg$ and 0.79$\arcdeg$, respectively, with TRGB distances
 of $3.94 \pm 0.27$~Mpc and $3.87 \pm 0.27$~Mpc \citep{kar02}. Both of these dSph objects have not been detected in HI.
  The position and TRGB 
 distance \citep[$3.98 \pm 0.29$~Mpc;][]{kar02} of the dwarf 
 irregular galaxy AM1318-444 (KK\,196) suggests it is in or beyond the 
 southern lobe of Cen\,A. The possibility of it having passed through the 
 radio lobe, causing any of the observed filaments in the area \citep{fea11} is investigated and dismissed as unlikely by \citet{wyk14}. 

\begin{table*}
\centering
\caption{Global Parameters for NGC 5128 and ESO 324-G024}
  \begin{tabular}{lccr}
  \hline
\large Parameter & \large NGC 5128 & \large ESO324-G024&\large Ref\\
   \hline
Other Names & Centaurus A, HIPASS J1324-42, & PGC 047171, HIPASS J1327-41,  & 1\\
&AM 1322-424, ESO 270-IG009  &AM 1324-411, UKS 1324-412&\\
D $\pm\ \sigma_{\rm D}$ (Mpc) & 3.8 $\pm$ 0.1 & 3.73 $\pm$ 0.43 & 2, 3\\
Optical Centre (RA, DEC) (J2000) & (13$^{\rm h}$25$^{\rm m}$28.9$^{\rm s}$, $-$43$\arcdeg$01$\arcmin$00$\arcsec$)& (13$^{\rm h}$27$^{\rm m}$37.4$^{\rm s}$, $-$41$\arcdeg$28$\arcmin$50$\arcsec$) & 4\\
Optical major-to-minor axis ratio ($b/a$) &\nodata&0.68&*\\
Inclination, $i$, of stellar disk  & \nodata &53$\fdg3$& *\\
$M_{\rm B}$ &$-$20.48&$-$15.39&3\\
$V_{\rm sys}$ (km s$^{-1}$)  &556 $\pm$ 10 &  528.1 &4, * \\
$V_{\rm LG}$ (km s$^{-1}$)  & 338 &282  & 4, *\\
$F_{\rm HI}$ (Jy km s$^{-1}$) &91.8 $\pm$ 13.2 &47.5 $\pm$ 5.5 &4 \\
Total \hi\ mass (M$_{\sun}$) & 4.37 $\times$ 10$^8$& 1.30 $\times$ 10$^8$&4\\
12 $+$ log (O/H) & 8.52 $\pm$ 0.03 &7.94 $\pm$ 0.11 &5,6\\
Total stellar mass (M$_{\sun}$) & 2.2($\pm$0.4) $\times$ 10$^{11}$& 1.7 $\times$ 10$^8$& 7, *\\
log $L_{\rm H\alpha}$ (erg s$^{-1}$) &\nodata&39.22&8\\
FUV (mag) &\nodata&$-$14.2 $\pm$ 0.02&*\\
WISE 22$\mu$m (mag) &\nodata& $-$9.031 $\pm$ 0.391&*\\
\hline
\tablerefs{*Indicates this work.;
(1) NASA Extragalactic Database; 
(2) \citet{har10}; 
(3) \citet{kar02};
(4) from HIPASS survey, \citet{kor04}; 
(5) \citet{wal12};
(6) \citet{lee07};
(7) \citet[][assuming a Salpeter IMF]{dea12};
(8) \citet{ken08} 
}\label{tab:1}
\end{tabular}
\end{table*}

\begin{figure*}
\begin{minipage}{0.55\textwidth}
{\includegraphics[scale=.28]{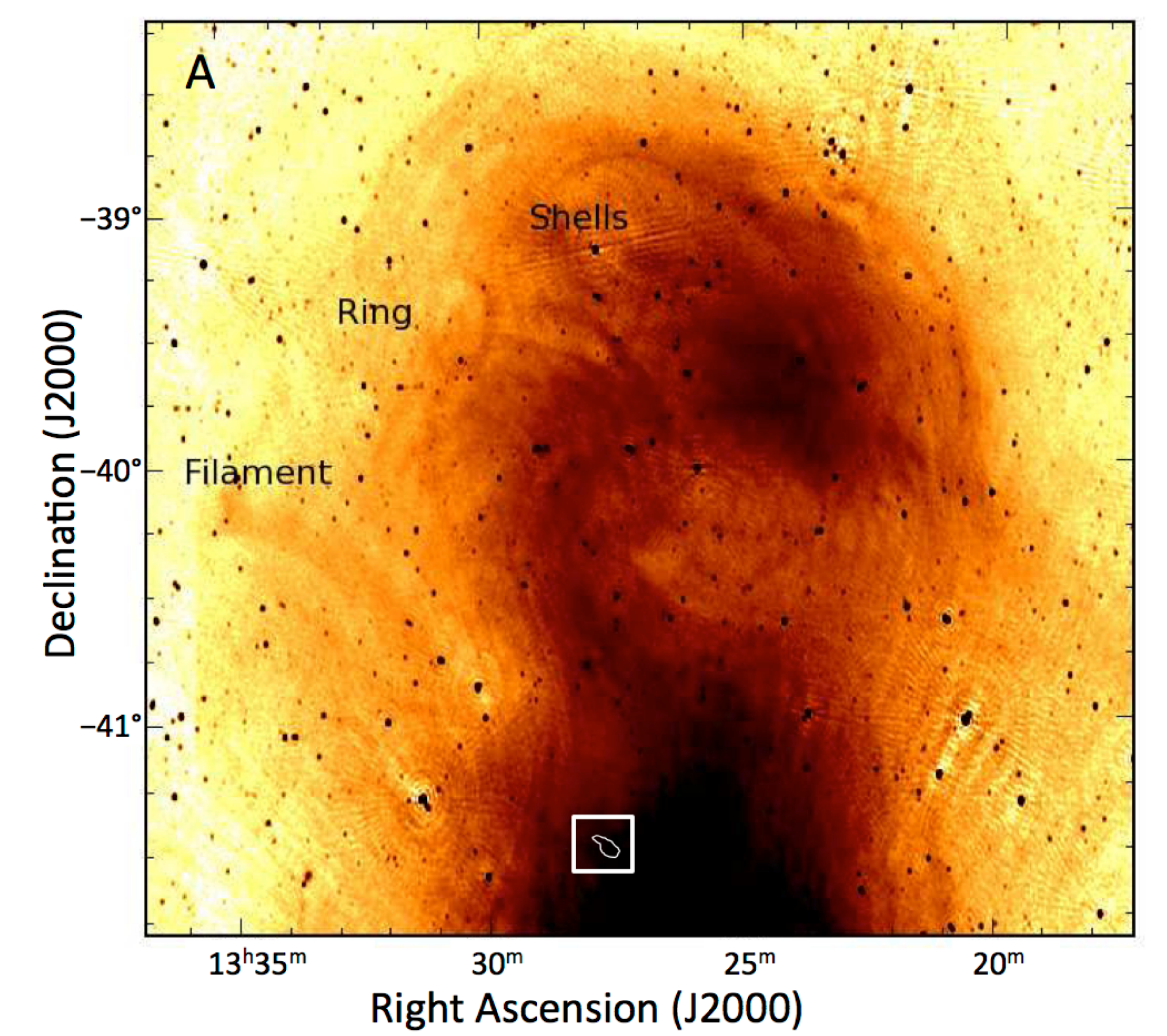}\label{fig:1b}} 
\end{minipage}
\hfil
\begin{tabular}{lc}
&{\includegraphics[scale=0.16]{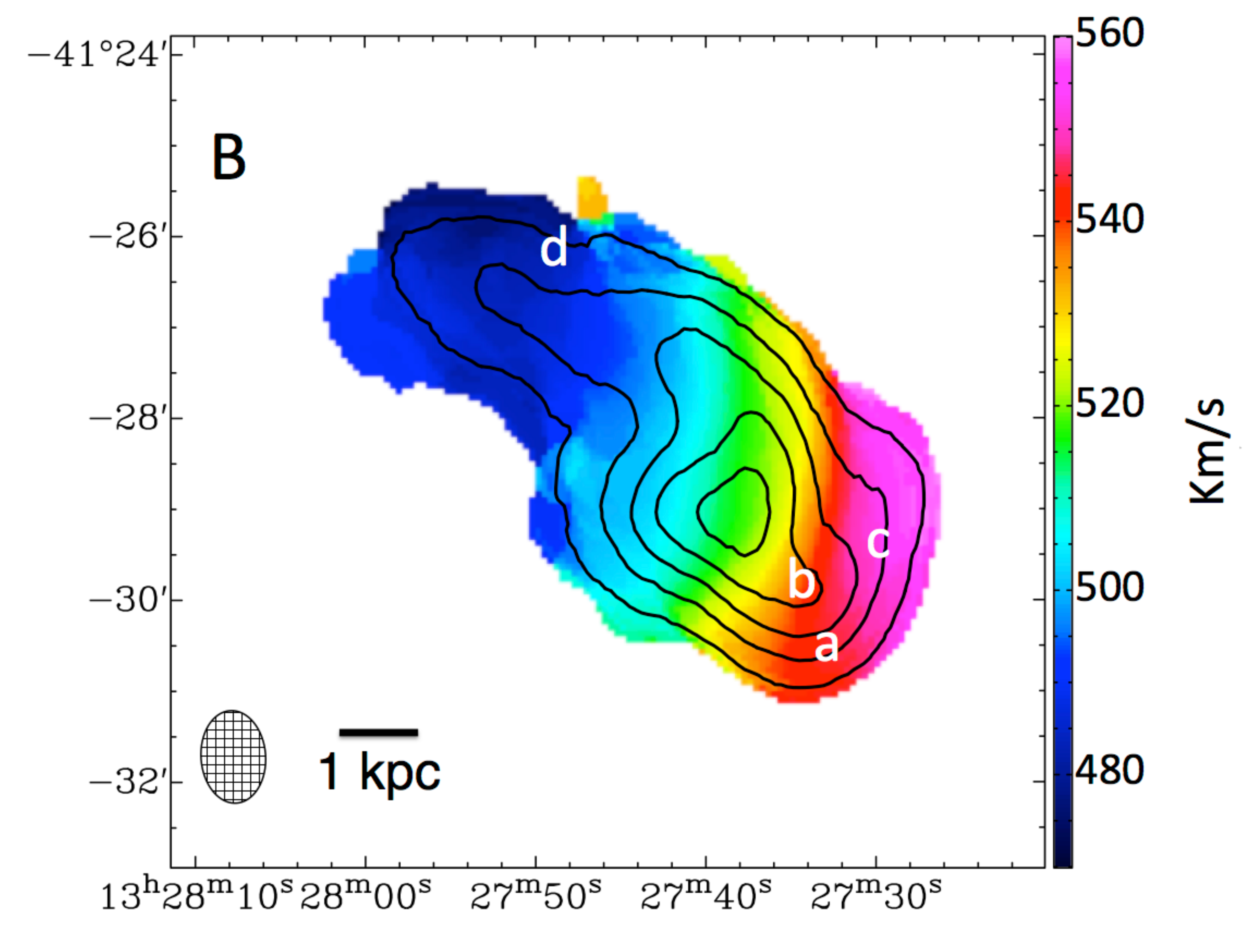}\label{fig:2b}}\\
&{\includegraphics[scale=.14]{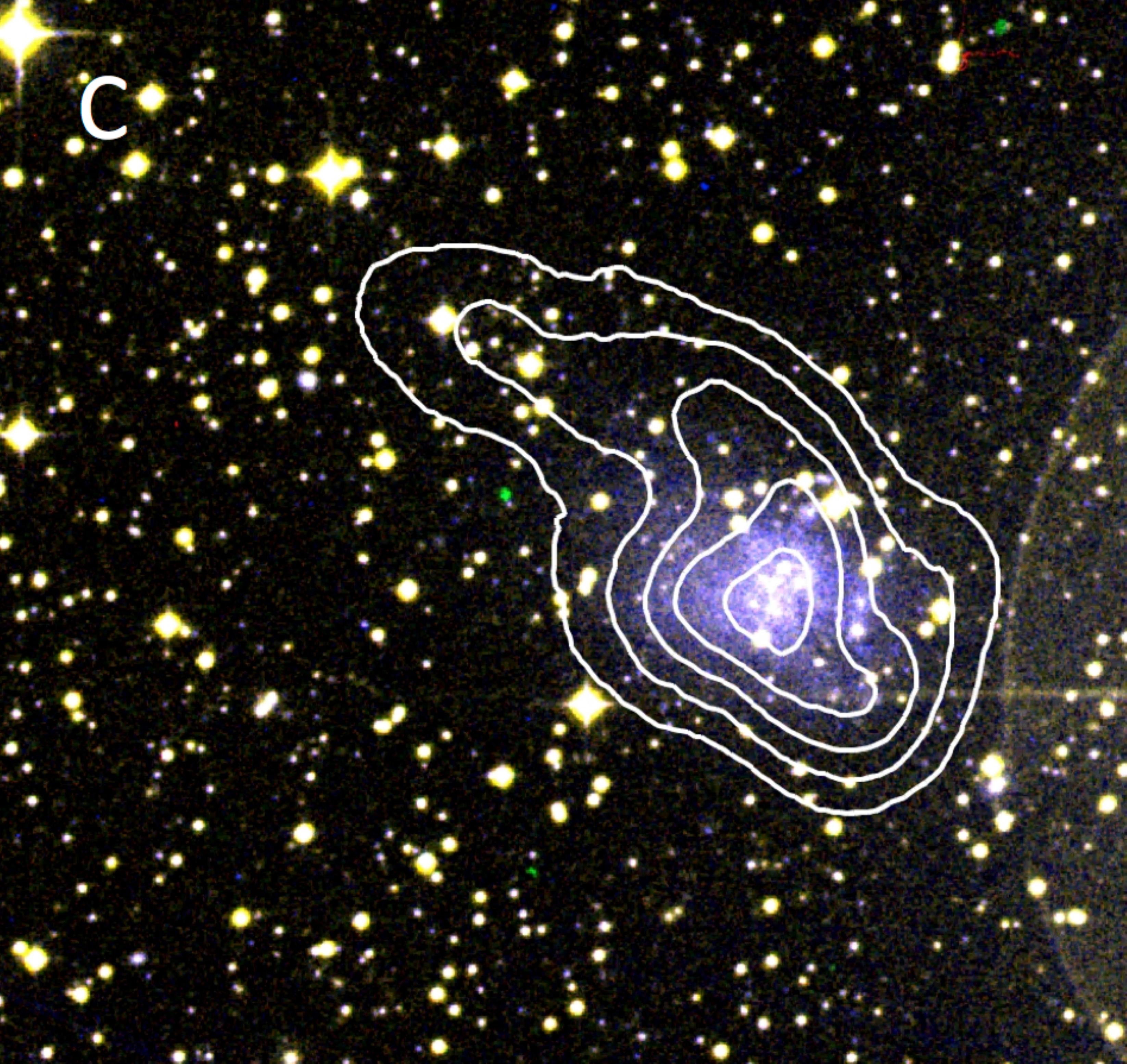}\label{fig:3d}}\\
\end{tabular}
\caption{(a) High spatial resolution ATCA + Parkes 1.4 GHz radio continuum image from \citet{fea11} with 
    three features (filament, ring and shells) marked. Overlaid is the outermost \hi\ contour
    of ESO324-G024 (to scale) in white, surrounded by a white box which corresponds to the image
    in panels (b) and (c).
    (b) Integrated \hi\ intensity contours overlaid on intensity-weighted integrated \hi\ velocity field.  (c) Optical RGB composite image made from the SuperCOSMOS $B_{\rm J}-$ and $R-$bands and the DSS\protect\footnotemark\ $I-$band image with \hi\ intensity contours overlaid. The \hi\ intensity contours in panels (b) and (c) are at \hi\ column densities $N_{\rm HI}$ = 2, 6, 10, 14, and 18 $\times$ 10$^{20}$ cm$^{-2}$. The resolution of the \hi\ data for ESO 324-G024 is shown by the oval in the bottom left corner of panel (b) and the letters a, b, c, and d demarcate the locations where line profiles were derived (see Section \ref{sec:2Dkin} for details). \label{fig:2}}
\end{figure*}
\footnotetext{The Digitized Sky Surveys (DSS) were produced at the Space Telescope Science Institute under U.S. Government grant NAG W-2166. The image used in this work is based on photographic data obtained using the UK Schmidt Telescope. The plates were processed into the present compressed digital form with the permission of this institution. The UK Schmidt Telescope was operated by the Royal Observatory Edinburgh, with funding from the UK Science and Engineering Research Council (later the UK Particle Physics and Astronomy Research Council), until 1988 June, and thereafter by the Anglo-Australian Observatory. }

Our paper is organized as follows: Section \ref{sec:data} describes the data that we will use in our investigation, Section \ref{sec:geo} explores the geometry of Cen A's radio lobe with respect to ESO 324-G024, Section \ref{sec:kin} dissects the gas kinematics of ESO 324-G024,  Section \ref{sec:discuss} contains the discussion, and Section \ref{sec:sum} concludes our paper with a summary.

\section{ESO 324-G024 Data and Analysis}\label{sec:data}
To investigate the dwarf galaxy ESO 324-G024 and its surroundings, we use a 
  suite of multi-wavelength data, including \hi\ spectral line data from the 
ATCA, as well as Parkes and ATCA 1.4 GHz 
  radio continuum and polarization data of Cen\,A.




	\subsection{\hi\ Spectral Line Data}\label{sec:hidata}


ATCA 21 cm observations were obtained in September 1998 in three arrays (375, 750D and 1.5D) 
  with on-source integration times of 600, 420, and 560 minutes, respectively.
  The 8~MHz bandwidth was divided into 512 channels, giving a channel width of 
  3.3 \kms\ and a velocity resolution of 4 \kms. The data are part of the 
  `Local Volume \hi\ Survey' (LVHIS) project \citep{kor08}. The Natural-weighted \hi\ cubes have a resolution of $66\farcs5 \times 43\farcs3$ and
  an rms of 3.5 mJy\,beam$^{-1}$. 

  We measure an integrated \hi\ flux density of 51.3 Jy\kms, in agreement with 
  the HIPASS value \citep{kor04}. Assuming $D$ = 3.73~Mpc, we derive
  an \hi\ mass of M$_{\rm HI}$ = $1.3 \times 10^8$ \msun; scaling by 1.4 to account for helium and metals \citep[][and references therein]{oh08} gives a total gas mass of 1.8 $\times$ 10$^8$ \msun. 
  The molecular gas content in ESO 324-G042 is expected to be minimal \citep[see e.g.,][]{ler06, buy06}. \color{black}

Figure \ref{fig:2} shows an \hi\ intensity contour of ESO 324-G024 
	overlaid onto the high spatial resolution ATCA $+$ Parkes radio continuum image of Cen A from \citet{fea11}.  There are some residual imaging artifacts present in the ATCA $+$ Parkes image, but, the detailed structure of the northern lobe and the true scale size of the \hi\ distribution of ESO 324-G024 is visible.
Figure \ref{fig:2} also shows integrated \hi\ intensity contours that depict a striking \hi\ tail that extends roughly 3.5 kpc in projection from the main body of the disk.  
 To better see the morphology of the \hi\ features in ESO 324-G024, we examine individual channel maps shown in Figure \ref{fig:chan}. The \hi\ tail is prevalent in nearly every channel.
 
\begin{figure*}
\centering
{\includegraphics[scale=.9]{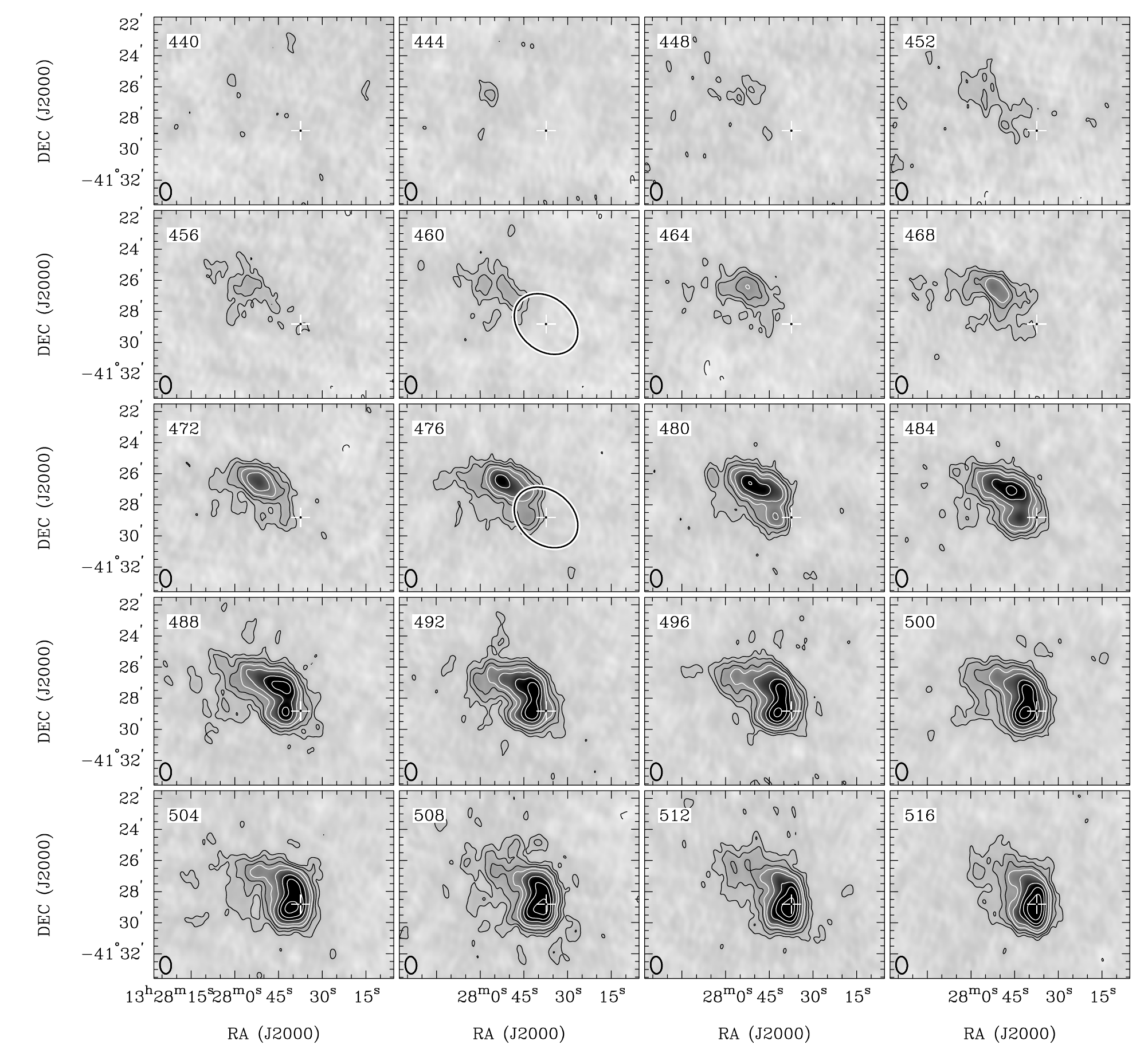}}
\caption{\hi\ channel maps  of ESO 324-G024 from the \hi\ data cube.  The velocity of each channel, in \kms, is given in the upper left corner and the optical center is marked with a cross. The contours are at \hi\ intensities of 5.4, 9, 14.4, 20, 30, 50, 70, and 100 mJy bm$^{-1}$.  The black oval in channels 460, 476, 544, and 560 marks the optical $B-$band 
radius at 25 mag $\rm arcsec^{-2}$.}
\end{figure*}
\addtocounter{figure}{-1}
\begin{figure*}
{\includegraphics[scale=.9]{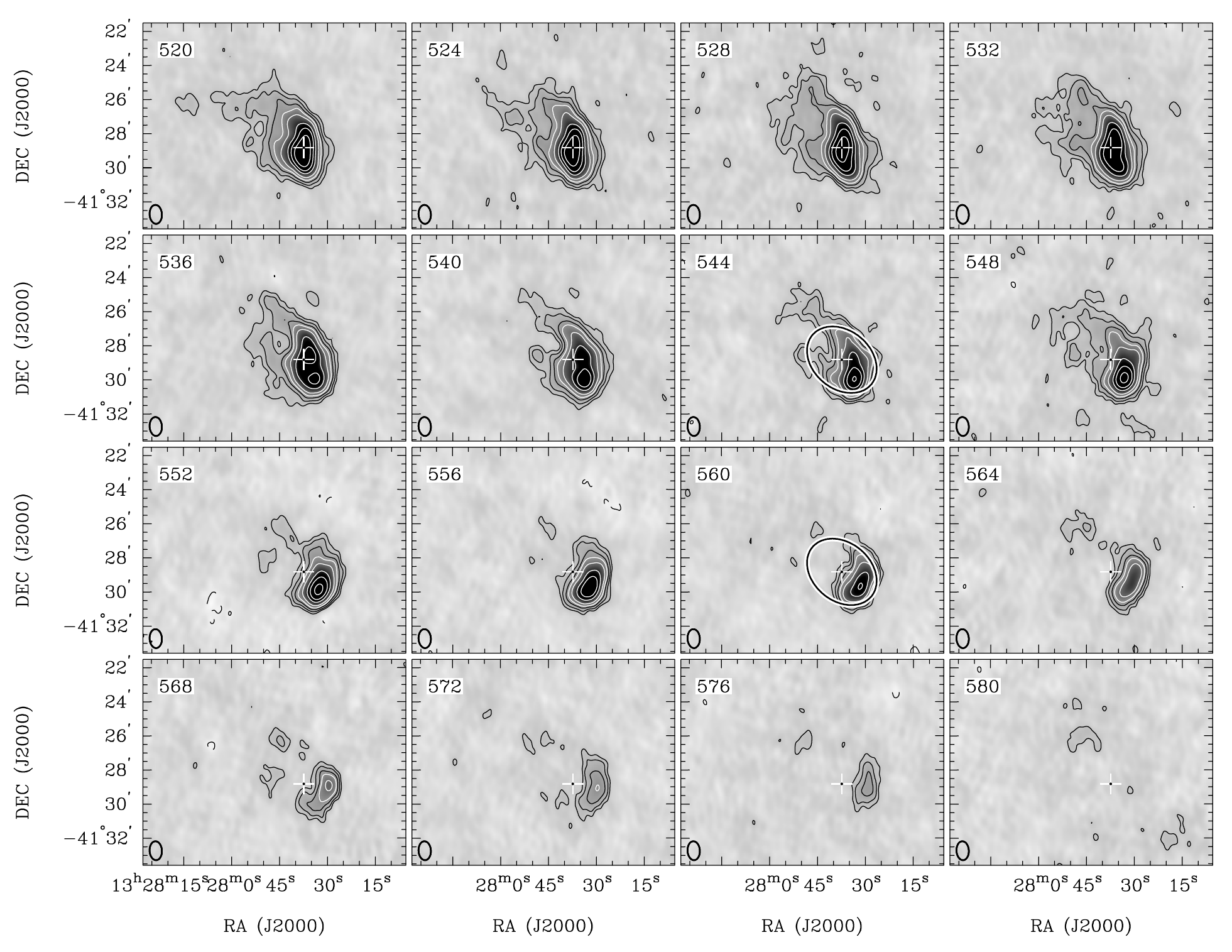}}
\caption{Figure continued from previous page}
\label{fig:chan}
\end{figure*}

	\subsection{Ancillary Data}\label{sec:optdata}

\begin{figure*}
\centering
\includegraphics[scale=.5]{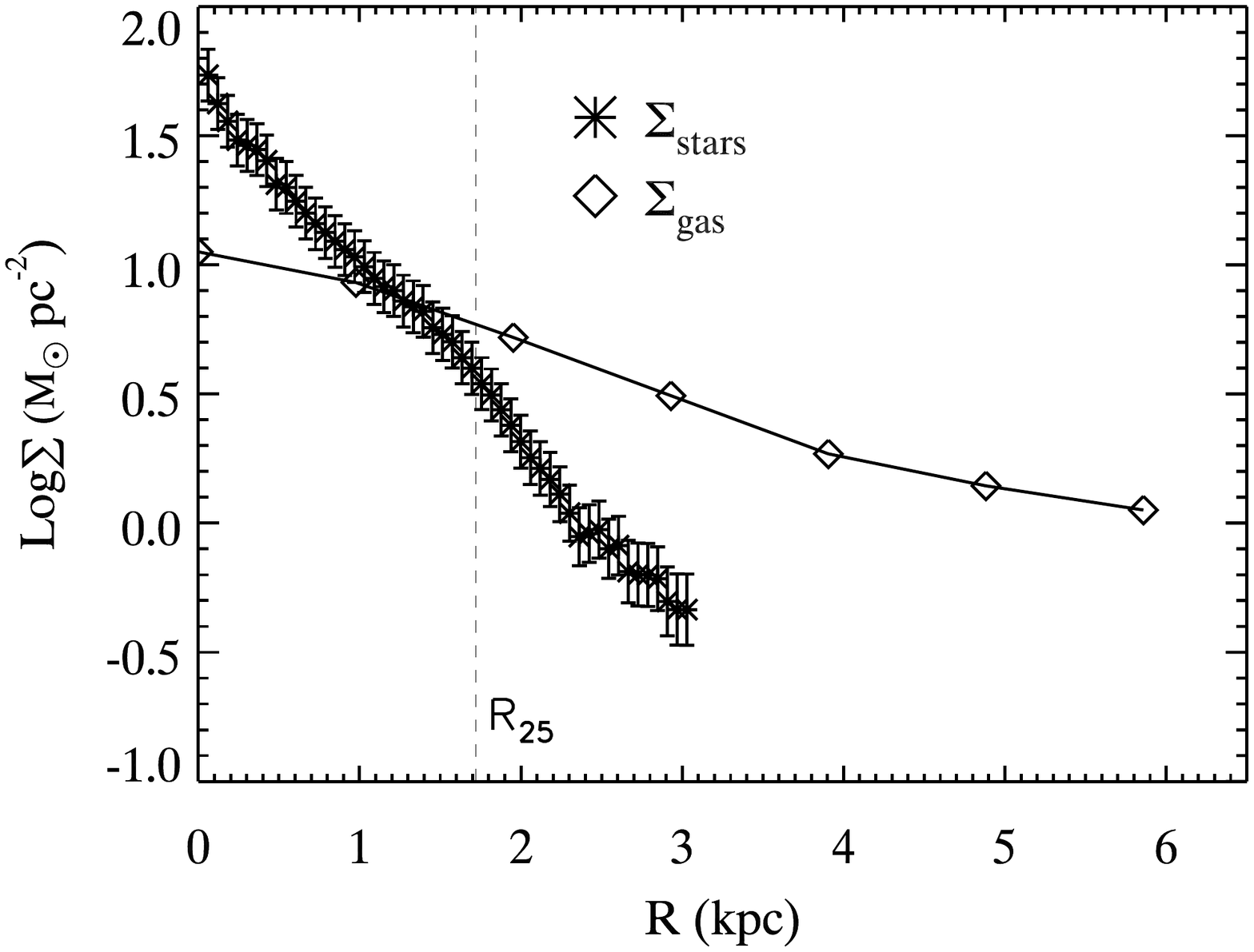}
\caption{Stellar and gas mass surface density profiles for ESO 324-G024. The optical $B-$band radius at 25 mag arcsec$^{-2}$, R$_{\rm 25}$, is the vertical dashed line marked for reference. }
\label{fig:4}\label{surfden}
\end{figure*}

We study the stellar component in ESO 324-G024 using \emph{B$_{\rm J}-$}band optical photometry from the SuperCOSMOS sky survey \citep{ham01a}, \emph {GALEX} ultraviolet data \citep{lee11}, \ha\ photometry \citep{lee09}, and \emph {Wide-field Infrared Survey Explorer (WISE)} 22$\mu$m data. 
 Figure \ref{fig:4} shows the 
 $B_{\rm J}-$band surface brightness profile from the SuperCOSMOS survey \citep{ham01a}, which we converted into a  mass surface density profile derived from the $B_{\rm J}-$band surface brightness profile which is measured from images from the SuperCOSMOS survey \citep{ham01a}
 using the mass-to-stellar light ratio method from \citet{bel03}.  We used a global $B-R$ color taken from \citet{lau89}
rather than direct measurements from the SuperCOSMOS images because of the color shifts of the SuperCOSMOS images as described in \citet{ham01b}. We determine an inclination of the stellar disk, $i = 53.3\arcdeg$ using the relationship, ${\rm cos}\ i = \frac{(b/a)^2 - q_0^2}{1-q_0^2}$, where $b/a$ is the minor-to-major axis ratio derived from the \emph{B$_{\rm J}-$}band photometry data and $q_0 = 0.4$ is the intrinsic disk thickness for dIrrs \citep{van88}. We derive a total stellar mass of 1.7 $\times$ 10$^8$ \msun\ for ESO 324-G024 and determine a half-light radius in $B-$band, i.e., effective radius, $R_{\rm e}$ = 830 $\pm$ 10 pc. 

ESO 324-G024 has a star formation rate (SFR) of 
1.6 \color{black} $\times$ 10$^{-2}$ M$_{\sun}$ yr$^{-1}$ 
corrected for internal dust attenuation \color{black} as determined by \citet{lee09} from the \ha\ imaging survey, 11HUGS \citep{ken08}. Comparatively, we determine an independent SFR following the method outlined by 
\citet{cal12} \color{black}by combining the \emph {GALEX} far-UV (FUV) data and the \emph{WISE} 22 $\mu$m data for ESO 324-G024. From this method, we determine
that ESO 324-G024 has a SFR of 1.14 ($\pm$0.17) $\times$ 10$^{-2}$ M$_{\sun}$ yr$^{-1}$, which is similar to the value derived from \ha\ alone.
We determine a star formation rate surface density $\Sigma_{\rm SFR}$ = 5 $-$ 9 $\times$ 10$^{-3}$ \msun\ yr$^{-1}$ kpc$^{-2}$.

	\section{Is ESO 324-G024 inside the northern radio lobe of NGC 5128?}\label{sec:geo}

ESO 324-G024 lies 
a projected 104 kpc from the core of NGC 5128 at a distance of 3.73 Mpc. 	
The northern radio lobe of NGC 5128 extends for $\sim$300 kpc in the north-south direction and roughly 200 kpc in the east-west direction \citep[][see Figure \ref{fig:lobe_orientation}]{sta13}.
However, the 3D volume and orientation of the lobes are not well known.  Based on the observed \hi\ absorption against the jet in the southern lobe and the lack of absorption in the northern lobe, \citet{str10} suggest that the northern inner lobe is pointed toward our line of sight while the southern inner lobe is pointed away.  
There have been studies on the inner jet angle to the line of sight. \citet{har11} find an angle of approximately 50$\arcdeg$ agrees with their models for the TeV $\gamma-$ray emission.  \citet{mul14} use high resolution VLBI data and find for the inner subparsec scale, a jet angle to the line of sight somewhere between 12$\arcdeg$ $-$ 45$\arcdeg$ from estimating the jet-to-counter jet ratio.
Unfortunately, determining the exact angle at which the giant lobes are oriented in the plane of the sky is complex and difficult and thus, the 3D orientation and volume is still unknown.  

\begin{figure*}
\centering
\subfigure[1.4 GHz polarization intensity map \label{fig:10a}]{\includegraphics[scale=.33]{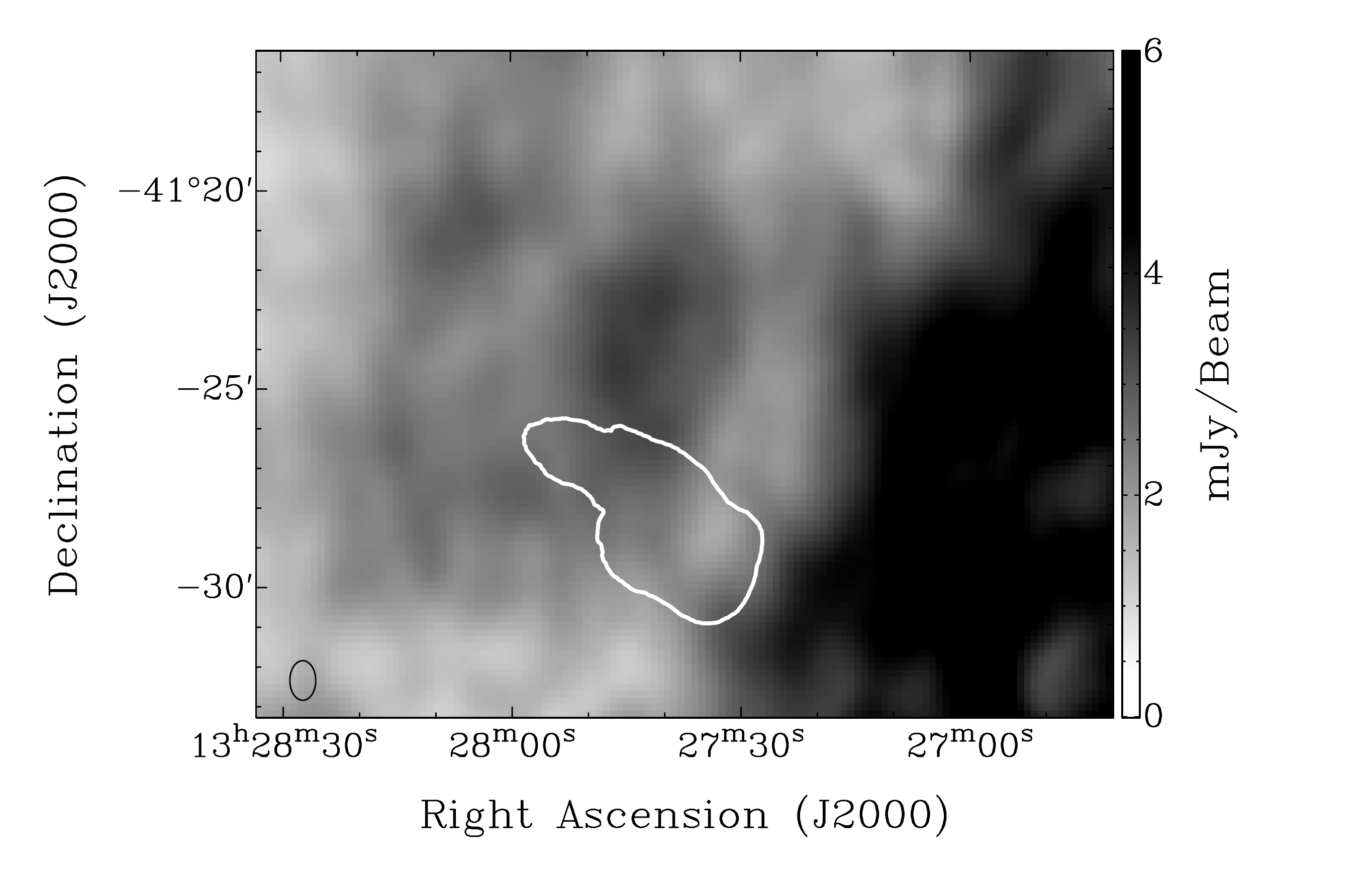}}
\subfigure[1.4 GHz rotation measure synthesis intensity map \label{fig:10c}]{\includegraphics[scale=.2]{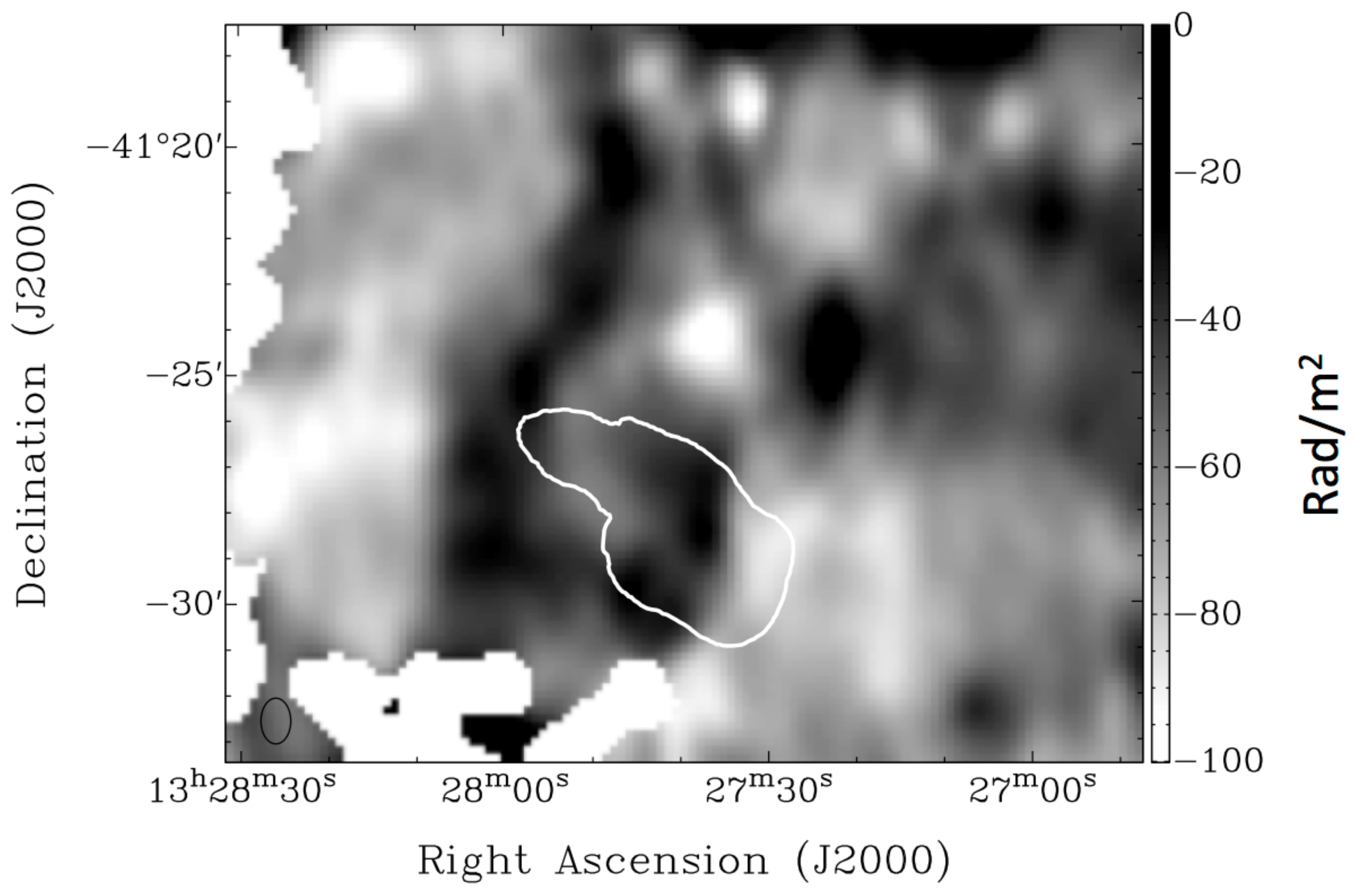}}
\subfigure[5 GHz polarization intensity map \label{fig:11a}]{\includegraphics[scale=.35]{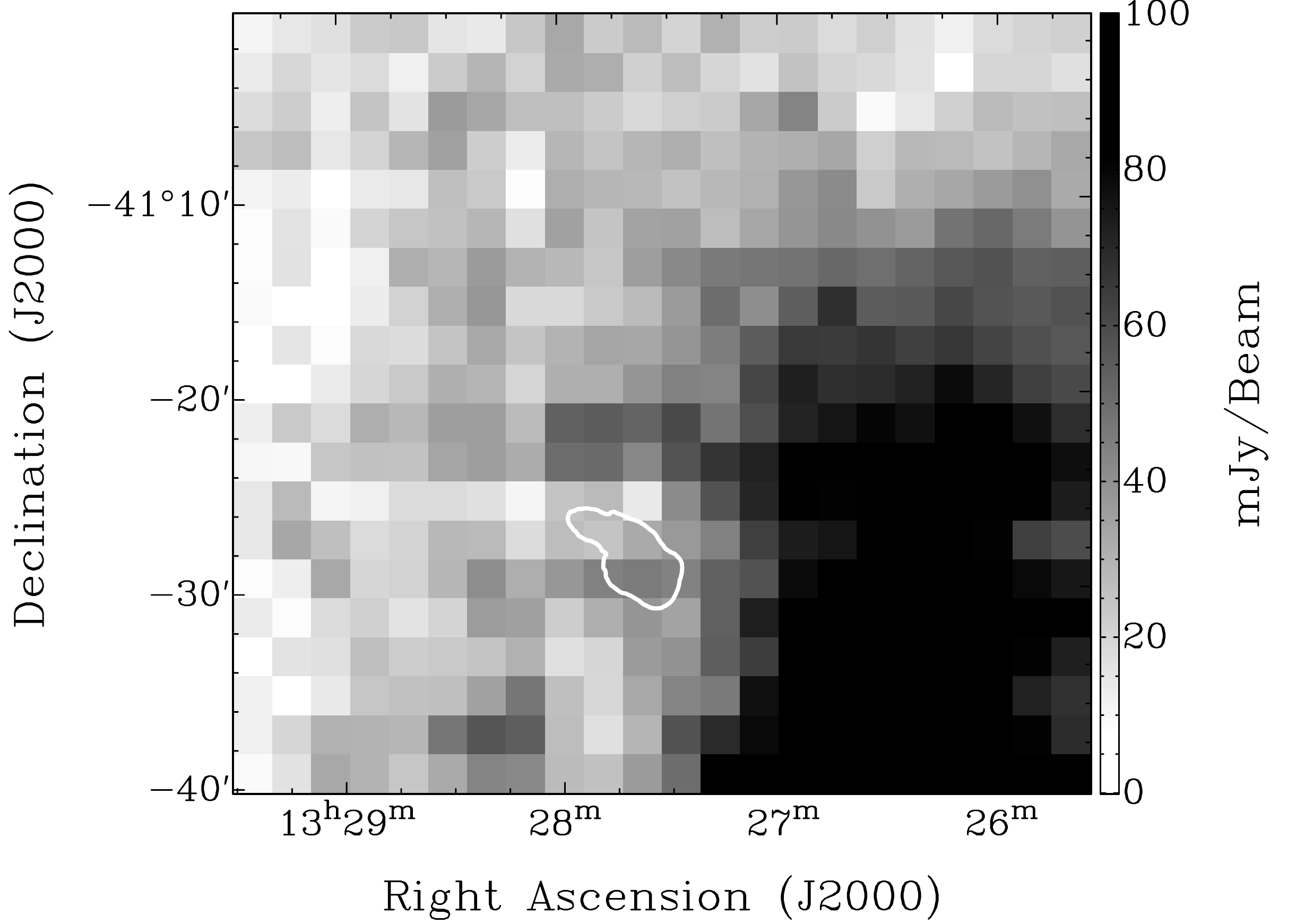}}
\subfigure[5 GHz fractional polarization intensity map \label{fig:11b}]{\includegraphics[scale=.21]{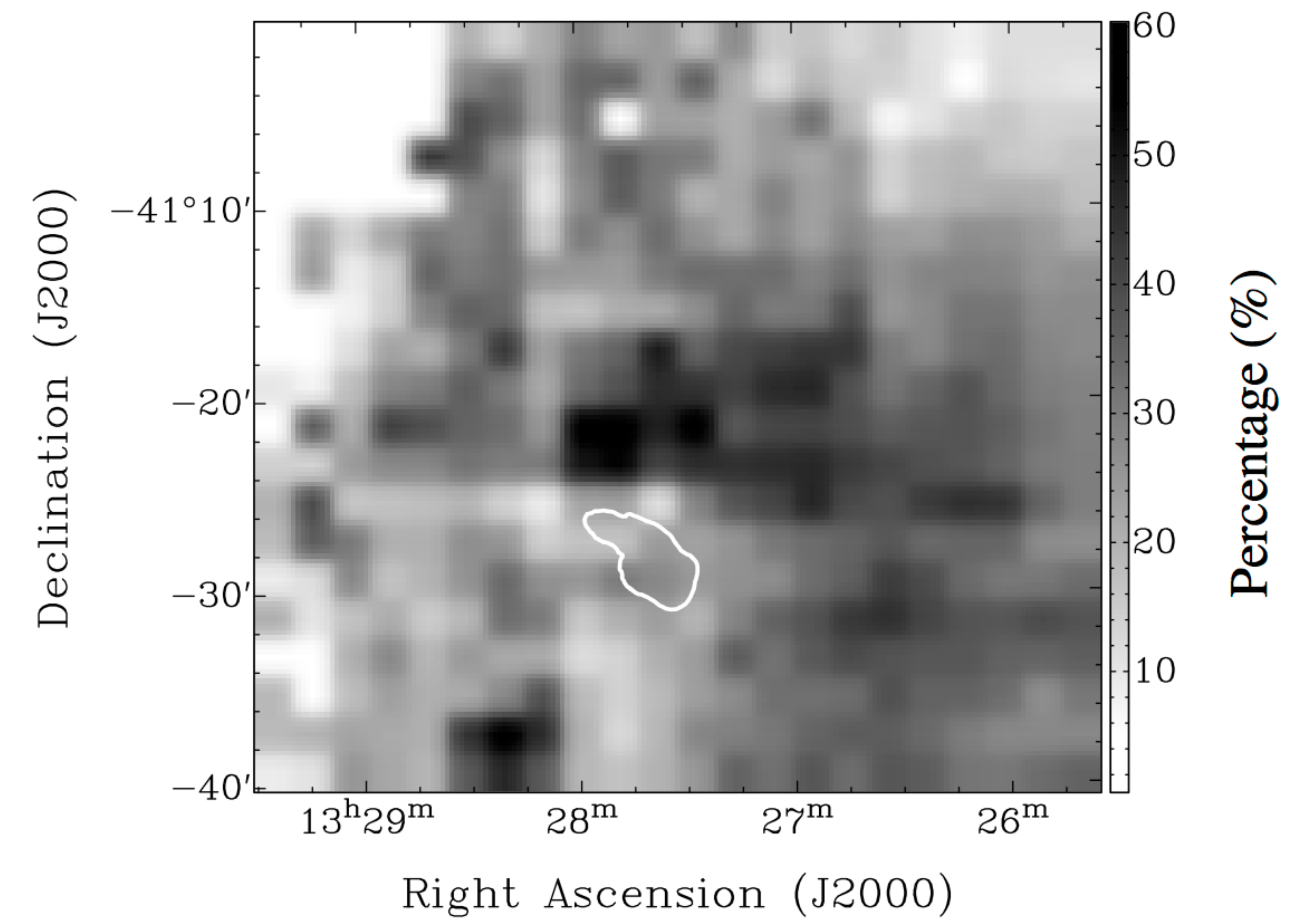}}
\caption{The top two panels are from \citet{osu13} combined with ATCA data and the bottom two panels are from \citet{jun93}.  The outermost HI contour of ESO324-G024 is shown in white as in Fig 2. We find no evidence for depolarization in the radio lobe of NGC 5128 at the location of ESO 324-G024 at either frequency.}
\label{fig:10}
\end{figure*}

To get a handle on whether ESO 324-G024 lies in front of, inside, or behind the northern radio lobe, we turn to the radio continuum data of Cen A.  
For our analysis, we study the Faraday rotation measure (RM), which is the rotation in the angle of polarization of linearly polarized electromagnetic radiation as it passes through a magnetized thermal plasma \citep{fea09}.  
Figure \ref{fig:10} shows the Parkes polarization intensity and RM maps from \citet{osu13} combined with ATCA data for a spatial resolution of about 1$\arcmin$ (1 kpc) with the outer \hi\ intensity contour of ESO 324-G024 overplotted for reference.  If ESO 324-G024 were in front of the lobe, then we would expect the background polarization emission from Cen A to be depolarized as it passes through the foreground dwarf.  However, 
at the location of ESO 324-G024, there is no depolarization signature in the lobe of Cen A as can be seen in Figure \ref{fig:10a} and there is no signal in the RM map (Figure \ref{fig:10c}).  Together, these results suggest that ESO 324-G024 is behind the radio lobe. 
As an analog to this scenario, \citet{fom89} show strong depolarization in the radio lobe of Fornax A at the location of NGC 1310, a spiral Sc galaxy, and the authors conclude that the only explanation for such depolarization must be that the galaxy is lying in front of the lobe. 
	
On the other hand, dwarf galaxies are known to have very low star formation rates compared to spiral galaxies \citep{hun04}, 
albeit comparable star formation efficiencies \citep{ler08, hua12}.  Therefore, it is plausible that ESO 324-G024 simply does not contain enough magneto-ionic material to cause a depolarization signal in the northern lobe of Cen A.  
We investigate if this is the case
by determining the minimum rotation measure dispersion, $\sigma_\mathrm{RM}$, required to show depolarization in the lobe. 
We would expect to see depolarization of the synchrotron emission from Cen A coincident with the galaxy if it were lying in front of or inside the lobe. Thus, if the polarized emission of the radio lobe is behind the dwarf, the formalism of \citet{bur66} applies, as we expect that turbulent cells in ESO 324-G024 are significantly smaller than the $\sim$1 $\mathrm{kpc}$ resolution of the continuum data\footnote{N.B., although the data shown in Figure \ref{fig:10} are ATCA $+$ Parkes data, the majority of the flux in the maps comes from the Parkes telescope, which has a beam size of 16 kpc at the distance of ESO 324-G024, which would only make this assumption more viable.}. In this case, the ratio of the observed polarized intensity to the polarized intensity expected in the absence of depolarization, known as the depolarization factor, is \citep{bur66}:
\begin{equation}
p \equiv \frac{L_\mathrm{observed}}{L_\mathrm{Cen A}} = \rm e^{(-2 \lambda^4 \sigma_\mathrm{RM}^2)}
\end{equation}
where $L$ is the linearly polarized intensity, 
$\lambda = \rm {wavelength} = 21.4 \, \mathrm{cm}$, and $\sigma_\mathrm{RM}$ the standard deviation of the rotation measures within the beam. From Figure \ref{fig:10a}, the average polarized intensity through the dwarf galaxy is $L_{\rm observed} \approx 2.4$ mJy bm$^{-1}$, while the average polarized intensity of the surrounding medium of Cen A near ESO 324-G024 is $L_{\rm Cen A} \approx 3.3$ mJy bm$^{-1}$.  Thus, $p$ > 0.7 and, therefore, if ESO 324-G024 were in front of Cen A, we would have $\sigma_\mathrm{RM}^2 < -\ln(p) / (2 \lambda^4)$, so $\sigma_\mathrm{RM} < 9 \, \mathrm{rad} \, \mathrm{m}^{-2}$.  As we discuss in the next section, this would be a very low value given the estimated random magnetic field and electron density in ESO 324-G024. We therefore conclude that it is unlikely that the dIrr is in front of the Cen A lobe.


			\subsection{Are there magnetic fields in ESO 324-G024?}
			
In order to determine the rotation measure dispersion for ESO 324-G024, we first attempt to estimate
the magnetic field strength in ESO 324-G024.
To date, only a handful of dwarf galaxies have detectable radio halos and magnetic fields \citep[see e.g.,][]{kep10, kep11, hee11, chy11, chy03, mao08, gae05}.  Nearly all of these objects are classified as starburst dwarf galaxies as defined by \citet{mcq10a,mcq10b}, who show that these dIrrs have sustained elevated SFRs for several hundred million years and that these rates of star formation are unsustainable over the lifetimes of the galaxies \citep{mcq10b}. 

ESO 324-G024 has an \ha\ SFR of 1.9 $\times$ 10$^{-2}$ M$_{\sun}$ yr$^{-1}$ (see Section \ref{sec:optdata}) and a total gas mass of 1.8 $\times$ 10$^8$ M$_{\sun}$ (see Section \ref{sec:hidata}), which produces an upper limit on the gas consumption timescale $\tau_{\rm gas}$ $\sim$ 13 Gyr, 
which is small compared to typical field dIrr galaxies \citep{hun12}.  \color{black}\citet{ken08} find an integrated \ha\ equivalent width of 44 $\pm$ 11 \AA\ for ESO 324-G024. 
We also examined the color magnitude diagram (CMD) of ESO 324-G024 from the publicly available Extragalactic Distance Database \citep{jac09} and find that ESO 324-G024 shows unambiguous signs of recent star formation based on the well populated blue helium burning branch and upper main sequence stars.  Because of its CMD and short gas consumption timescale, which suggests that its recent SFR is elevated, we find that 
ESO 324-G024 has properties consistent with known starburst dIrr galaxies \citep{mcq10a, mcq11, hun12}.
\color{black}Based on the $\Sigma_{\rm SFR}$ and magnetic field relationship identified by \citet[][see their Figure 7(a)]{chy11} for the Local Group dIrrs, we can crudely estimate a magnetic field strength for ESO 324-G024 in the neighborhood of $\sim$5 $\mu$G.  

We can 
roughly \color{black}estimate the ordered linear polarization in ESO 324-G024 by making some assumptions of its electron number density, $n_e$, and path length, $L$, through the disk.  From the three-dimensional kinematics \hi\ model (Section \ref{3Dmod}), we find that ESO 324-G024 has a disk scale height of 720 pc.  According to \citet{nic14}, isolated dIrr galaxies have $n_e$ values that range from 5 cm$^{-3}$ to $\sim$100 cm$^{-3}$, generally. For the path length, $L$, of ESO 324-G024, we use the approximate scale height determined from the three-dimensional modeling (see Section 
\ref{3Dmod}\color{black}), which is probably an underestimation of the path length since the path length is often roughly twice the scale height.  So, if we assume $n_e$ = 5 cm$^{-3}$ and $L$ = 720 pc we find an upper limit on the dispersion of the line-of-sight magnetic field in the galaxy to be,
\begin{equation}
\sigma_{B||} \le \frac{\sigma_{RM}} {n_e \times L} = 0.3\ \mu {\rm G}
\end{equation}
This is likely inconsistent with our estimated $5 \, \mu \mathrm{G}$ magnetic field in the galaxy.

The percentage of polarization in emission from gas with a magnetic field of 5 $\mu$G in ESO 324-G024 is difficult to determine, especially from this rough approximation.  \citet{vol08} show in their 6 cm radio continuum data of NGC 4501 that the polarized radio continuum emission is enhanced from ram pressure compressing the magnetic fields in the galaxy. 
The compression of magnetic fields from ram pressure was first proposed as an explanation for the displacement in the far-infrared$-$radio continuum correlation observed in cluster galaxies \citep{bos06}. \color{black}
If ESO 324-G024 is undergoing ram pressure forces, which is likely based on the morphology of the \hi\ gas, then one might expect similar compression of a magnetic field in ESO 324-G024 and
thus, ESO 324-G024 may also have an enhancement in polarized emission.  Therefore, it is possible that ESO 324-G024 has polarized emission, but the foreground emission from Cen A is overpowering that of the dwarf galaxy.

For comparison, \citet{gae05} find that the Large Magallenic Cloud (LMC) has a uniform magnetic field $B_0$ $\sim$ 1$\mu$G and a random field $B_{\rm R}$ $\sim$ 4.1$\mu$G, which gives a large-scale, total magnetic field strength $B_{\rm T}$ $\sim$ 4.3$\mu$G, which is similar to our estimated field strength of 5$\mu$G for ESO 324-G024.  The LMC is also a starburst dIrr galaxy in the throes of an interaction with our Milky Way galaxy and is similar to the interaction between ESO 324-G024 and NGC 5128, except for the giant radio lobe present in Cen A.   \citet{gae05} find a rotation measure dispersion $\sigma_\mathrm{RM} =$ 81 rad m$^{-2}$ for the LMC, which is significantly higher than our previously determined threshold of 9 rad m$^{-2}$ required for ESO 324-G024 to show depolarization in the lobe of NGC 5128 if it were in front of the lobe.  If the magnetic fields of ESO 324-G024 are compressed from ram pressure forces, then the rotation measure variance could be higher than that of the LMC, which is not in the presence of a giant radio lobe. 

As discussed in \citet{osu13}, it is expected that the majority of the observed polarized emission at 1.4 GHz from the lobe is dominated by that from the near side because there is internal Faraday rotation in the lobes.  Therefore, it may be possible that ESO 324-G024 is inside the lobe but near the back edge.  To see if this is the case, we examine the 5 GHz continuum data from \citet{jun93} since this frequency should be sensitive to the linearly polarized emission through approximately the entire depth of the lobes.  Figure \ref{fig:10} panels (c) and (d) show the 5 GHz continuum maps for NGC 5128 with the outer \hi\ contour of ESO 324-G024 overlaid for reference.  Although the quality of these data is poor, there is no clear evidence for depolarization at the location of ESO 324-G024.


We see no depolarization signal in the vicinity of ESO 324-G024 in either the 1.4 GHz or 5 GHz data for the northern radio lobe of Cen A and therefore, we conclude that it is most likely behind the lobe.  

\begin{figure*}
   \centering
   \includegraphics[scale=.35]{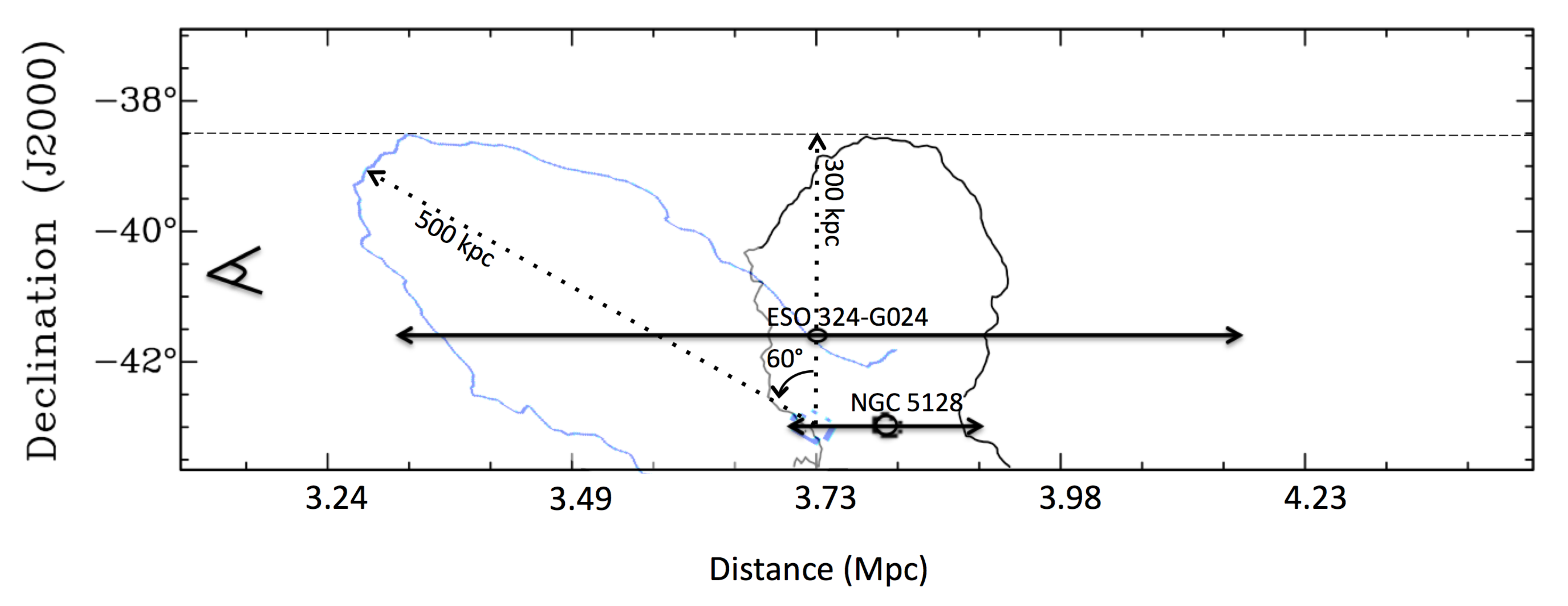} 
   \caption{Declination versus distance for NGC 5128 with the northern radio lobe and ESO 324-G024. We plotted a single contour at 3900 K of the continuum data in Figure \ref{fig:1a} (black contour) and made the assumption that the northern radio lobe is as deep as it is wide.  Our viewing angle is from the left as indicated by the eye in the plot. The solid arrow represents the distance uncertainty for NGC 5128 as determined from the tip of red giant branch from \citet{har10}.  The lobe shown at an angle of 60$\arcdeg$ (
   blue \color{black}contour) is to suggest one possible orientation for the northern lobe if NGC 5128 and ESO 324-G024 are at the same distance. The dashed line is to guide the eye to show that the lobe projected on the sky extends roughly $\sim$300 kpc in projection at 3.73 Mpc if it were in the plane of the sky and $\sim$500 kpc if it were inclined by 60$\arcdeg$.}
   \label{fig:lobe_orientation}
\end{figure*}

Figure \ref{fig:lobe_orientation} shows a schematic of the assumed distances to NGC 5128 and ESO 324-G024 versus declination.
This schematic shows the uncertainties in the distances to ESO 324-G024 and NGC 5128 (solid arrows).  The measured distances are consistent with ESO 324-G024 lying within the radio lobe if the lobe is in the plane of the sky.  
However, we have shown that ESO 324-G024 is behind the northern radio lobe and this puts some constraint on the orientation of the lobe and also on the distance to NGC 5128 and ESO 324-G024: the closer ESO 324-G024 is to the observer (relative to NGC 5128), the larger the required orientation angle. Two possible scenarios are shown in Figure \ref{fig:lobe_orientation}, one showing the northern lobe inclined by 60$\arcdeg$ that assumes ESO 324-G024 and NGC 5128 are at the same distance and that the lobe is as deep as it is wide.  
The second scenario is \color{black}the giant northern radio lobe of Cen A is in the plane of the 
sky. In order for ESO 324-G024 to be behind the lobe in this scenario, the distance to NGC 5128 much be less than 3.8 Mpc and/or the distance to ESO 324-G024 must be farther than 3.73 Mpc.  On the other hand, the lobes are relaxed structures \citep{osu13} so they are not necessarily symmetric. In this case, material could be at different distances along the line of sight.  As  \citet{str10} have already suggested that the northern inner lobe is closer than the southern inner lobe, it is probable that the northern lobe is inclined toward our line of sight such that ESO 324-G024 is behind it.  

\section{A Kinematic Perspective}\label{sec:kin}	

The distorted \hi\ morphology of ESO 324-G024 is striking and we investigate the kinematics of the gas to help better understand the mechanisms responsible for creating this morphology.  If the \hi\ tail  was formed from tidal interactions, then we would expect the gas kinematics to be distinct from the circular rotation in the disk \citep{bou04}.  On the other hand, if the \hi\ tail was formed from stripping via ram pressure forces, then the kinematics of the gas will likely resemble the circular rotation of the disk \citep{ken14, chu07}. To see if the \hi\ tail follows the motions in the disk, we model the kinematics from two different perspectives. First, we use 
a two-dimensional double Gaussian decomposition technique to separate bulk motions from non-circular motions. We then apply a two-dimensional tilted-ring model to the bulk velocity field to determine the rotation curve of the galaxy.  Second, we use a three-dimensional tilted-ring model that is applied directly to the \hi\ data cube and we compare the results of the two methods.

	\subsection{Two-dimensional Kinematic Model}\label{sec:2Dkin}
 		\subsubsection{\hi\ Double Gaussian Decomposition}

Tilted-ring modeling to derive rotation curves of disk galaxies has evolved greatly in recent times.  One of the most important improvements made to tilted-ring analysis for dwarf galaxies is extraction of reliable velocity fields over which the rings are parameterized. Dwarf galaxies have shallow gravitational potential wells, low shear, and thick disks all of which can make it easy for the gas within these systems to move about with motions counter to the overall rotation of the system. These non-circular motions cause asymmetries in an \hi\ line profile at various locations over the disk of a dwarf and they make it problematic to model the rotation curve from a standard intensity-weighted mean velocity field. Fitting a single Gaussian to these asymmetric \hi\ line profiles is also a poor way to create a velocity field when non-circular motions are present. Therefore, we employ a double Gaussian decomposition technique in order to separate bulk from non-circular motions. 
We briefly describe this method here and refer the reader to the works by \citet{oh08, oh11, oh14} for a full comprehensive description.  

\begin{figure*}
\centering
\subfigure[Bulk velocity field \label{fig:5a}]{\includegraphics[scale=.18]{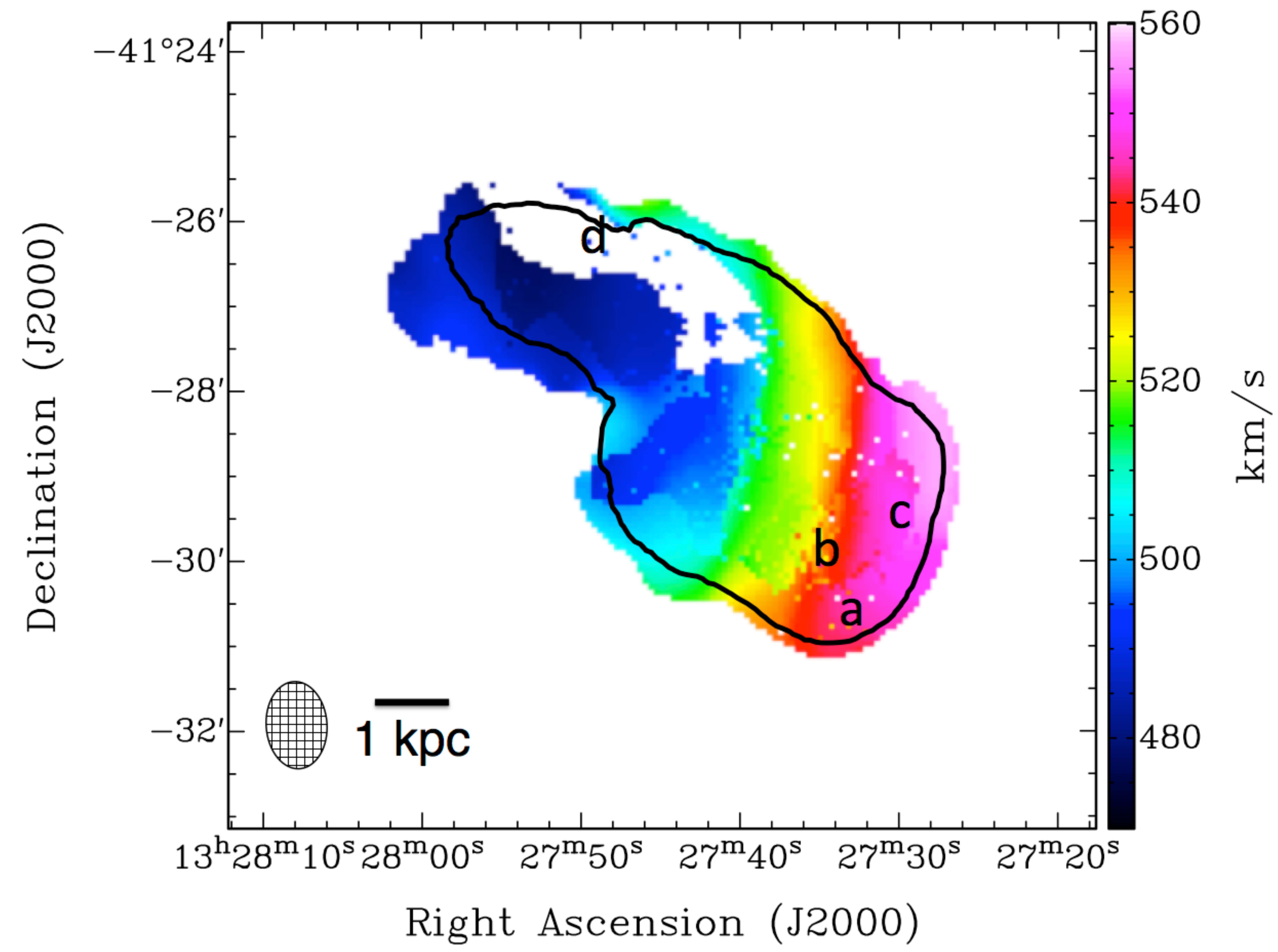}}
\subfigure[Model velocity field \label{fig:5d}]{\includegraphics[scale=.18]{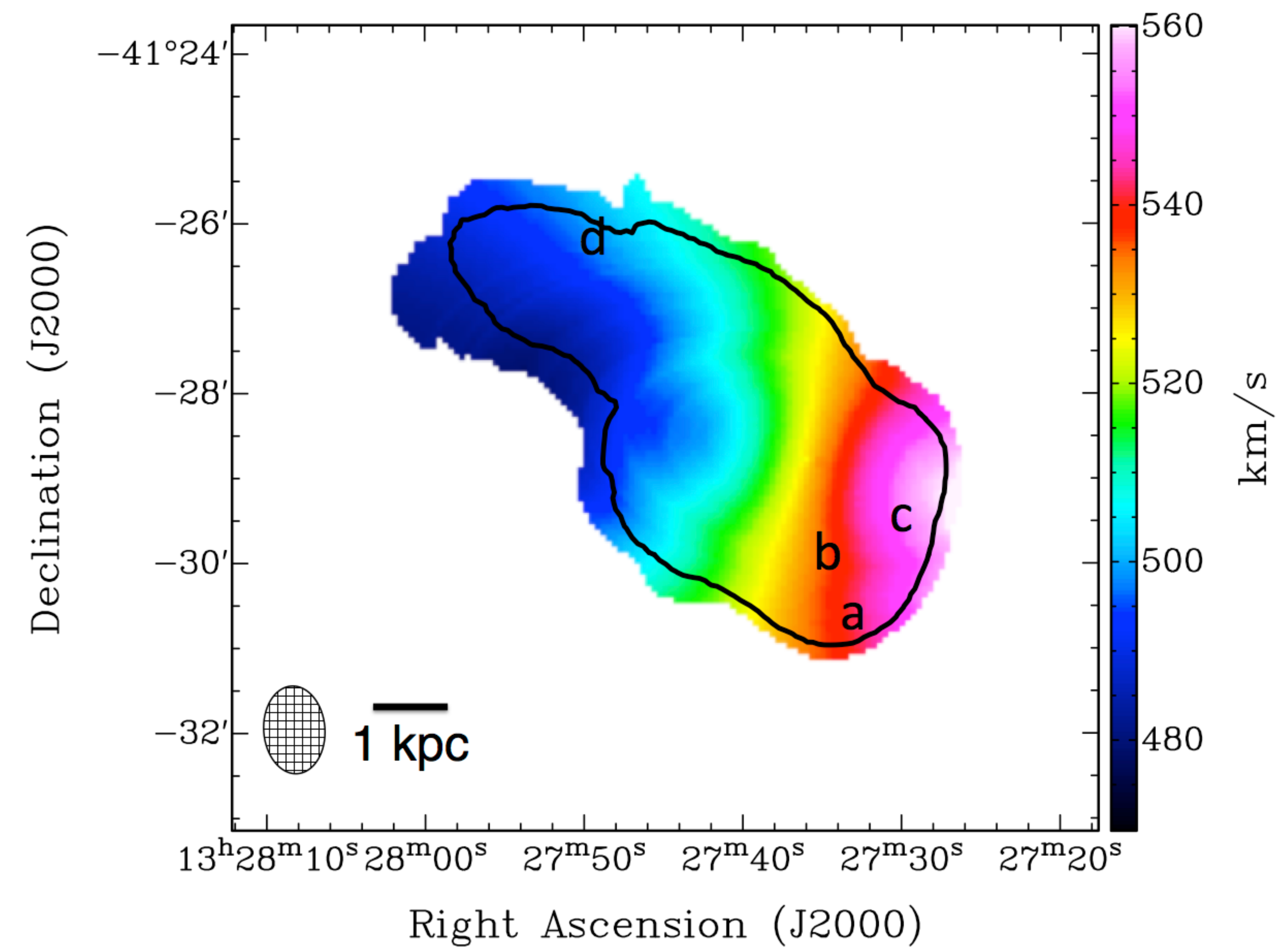}}
\subfigure[Strong non-circular motion velocity field \label{fig:5b}]{\includegraphics[scale=.18]{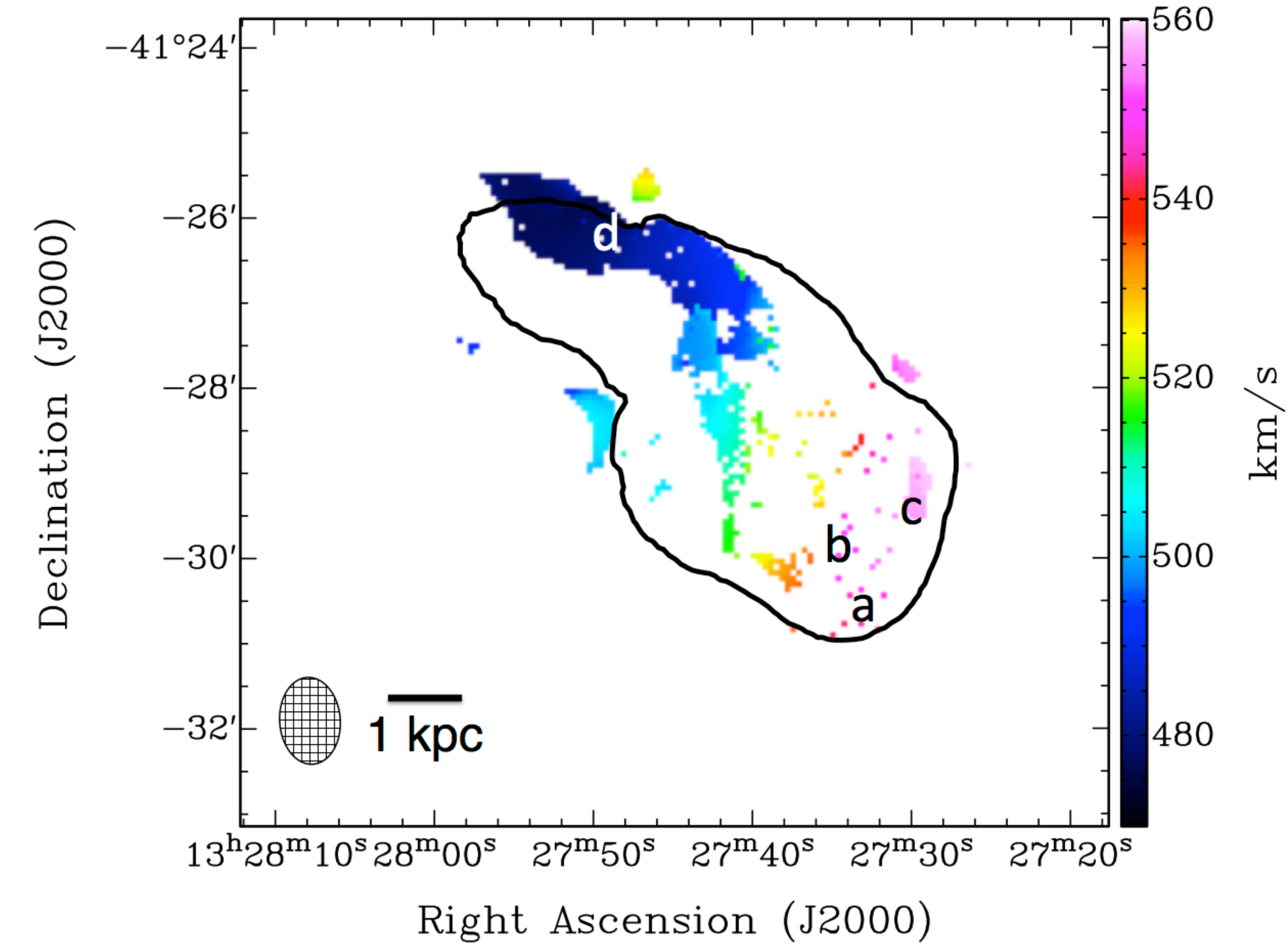}}
\subfigure[Weak non-circular motion velocity field \label{fig:5c}]{\includegraphics[scale=.18]{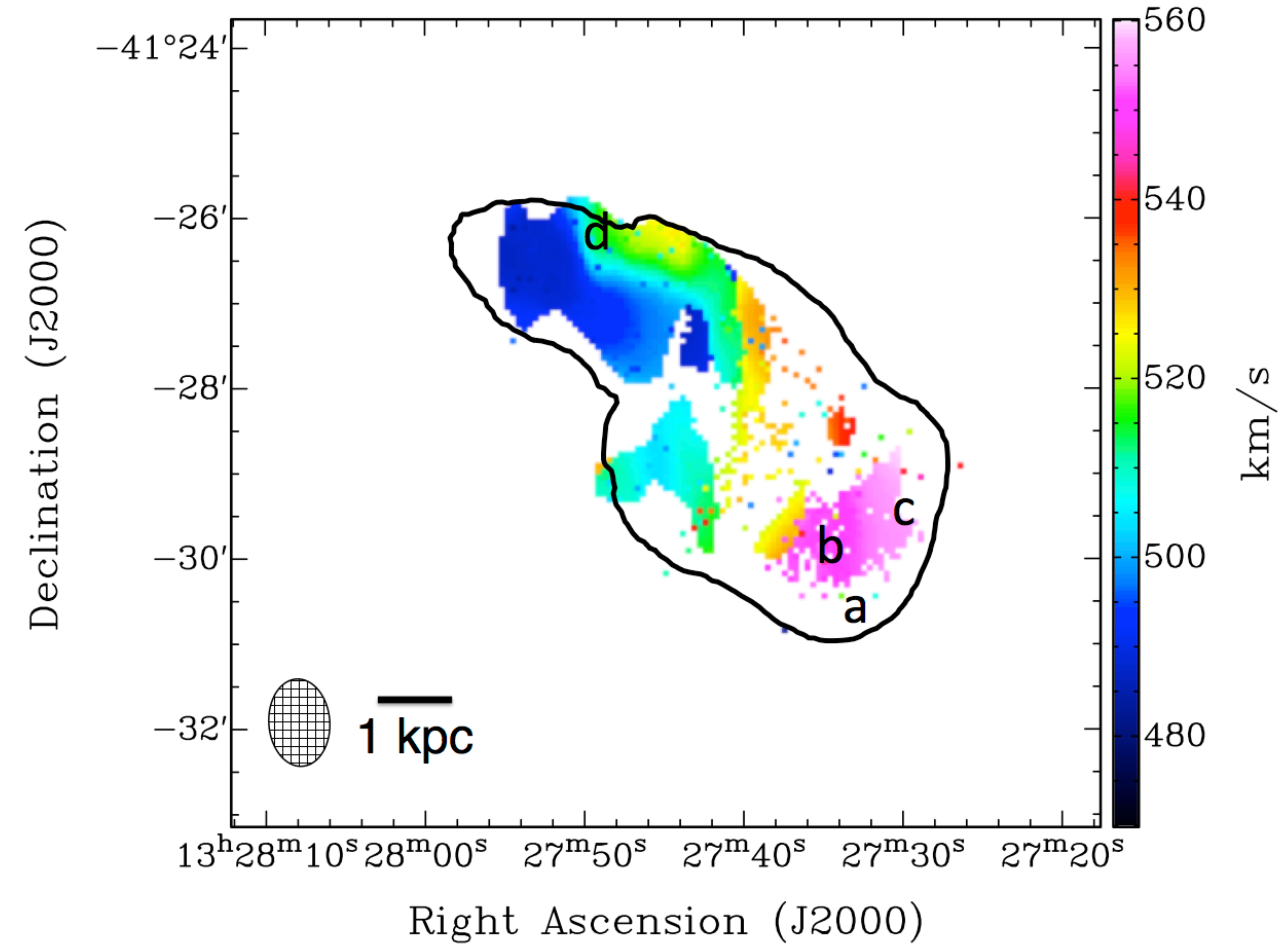}}
\caption{The results of the double Gaussian decomposition method for modeling the gas kinematics in ESO 324-G024. 
The outer \hi\ intensity contour from Figure \ref{fig:2} is overlaid for reference and is at $N_{\rm HI}$ = 2 $\times$ 10$^{20}$ cm$^{-2}$.  These velocity fields have been masked using the integrated intensity map from Figure \ref{fig:2} so that only real emission is used in the tilted-ring analysis. Locations marked `a', `b', `c', and `d' indicate the locations at which line profiles were extracted and shown in Figure \ref{fig:7}.  See text for more details.}
\label{fig:5}
\end{figure*}

\begin{figure*}
\centering
\includegraphics[scale=.2]{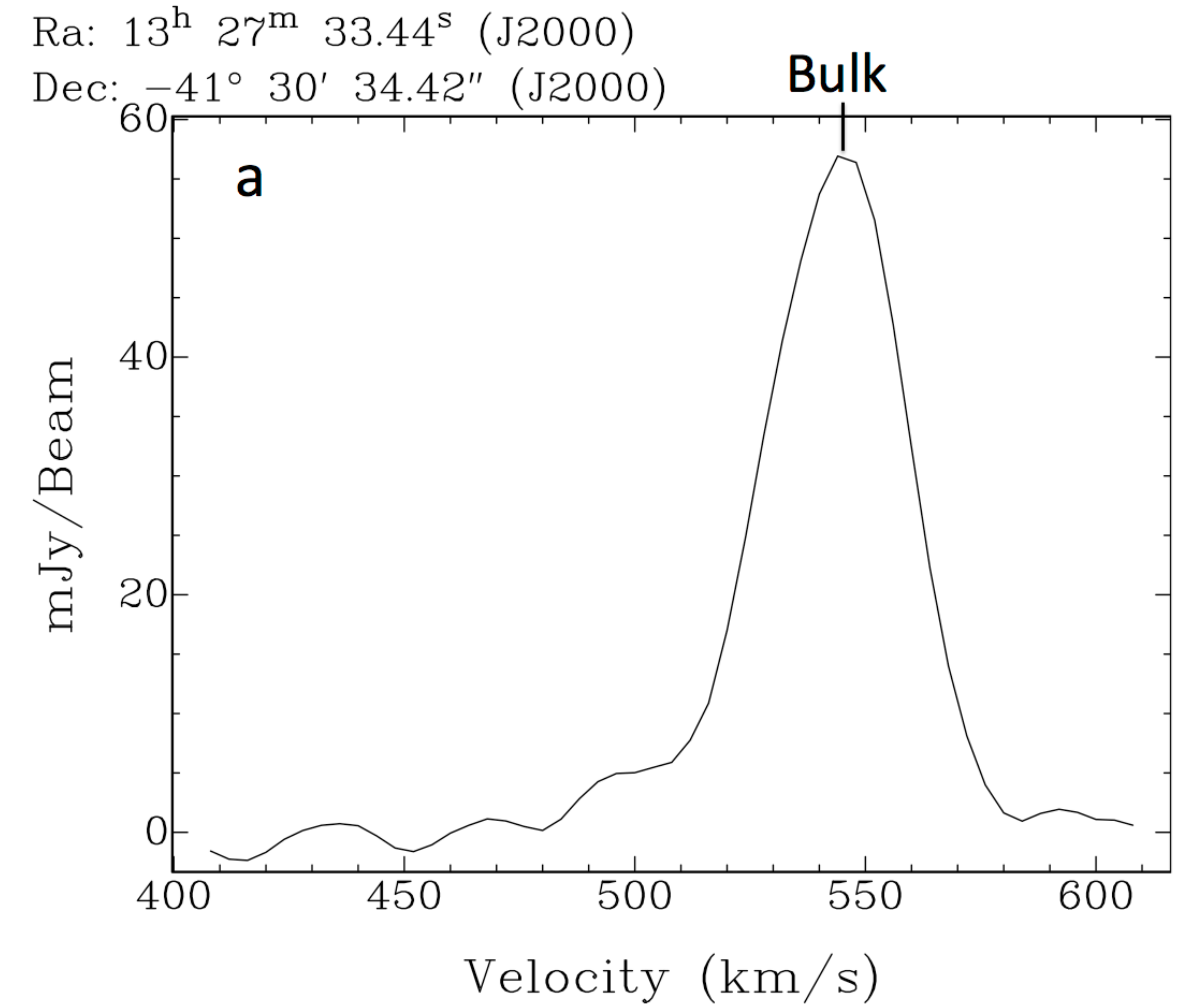}\label{fig:7a}
\hfil
\includegraphics[scale=.2]{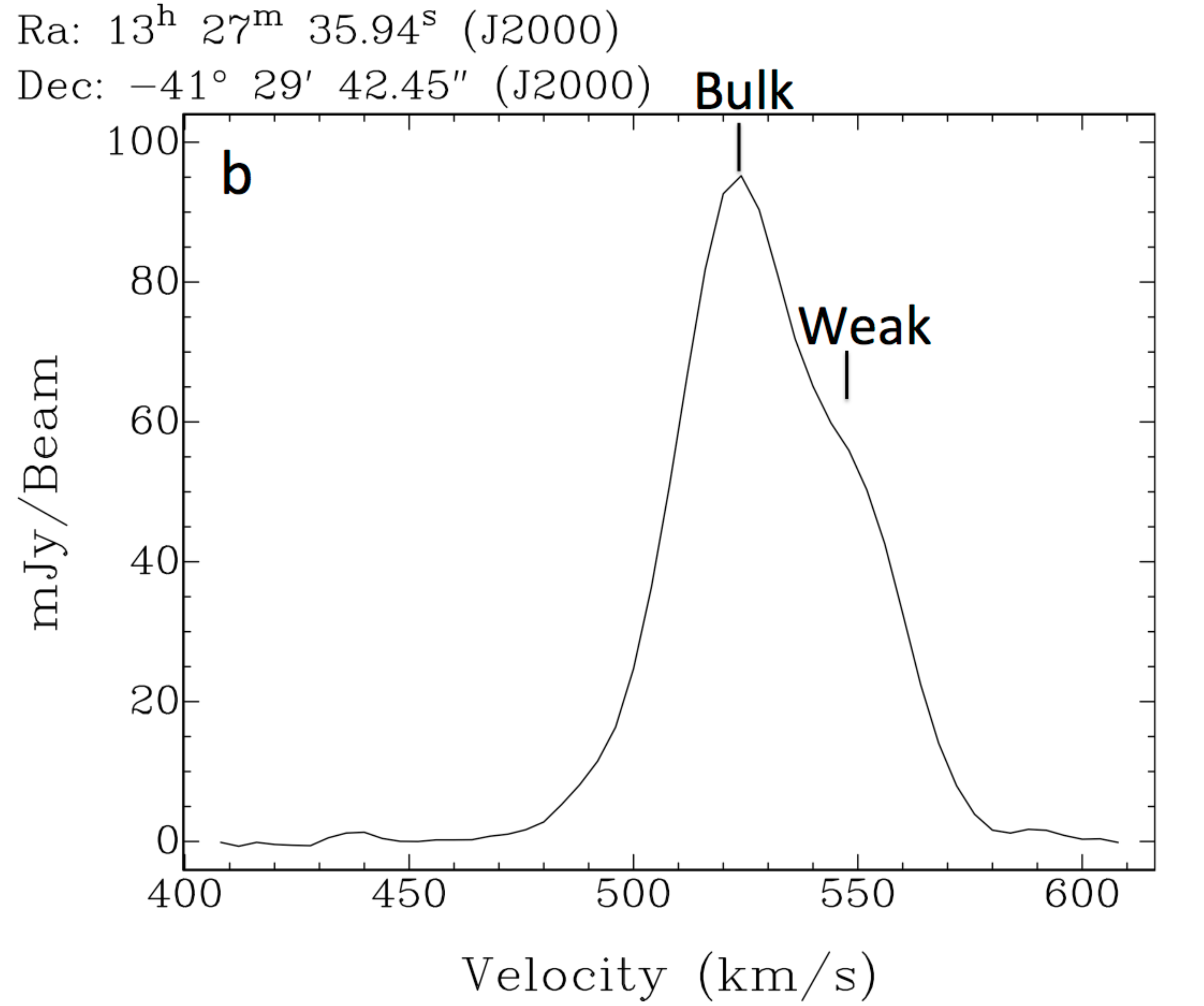}\label{fig:7b}
\hfil
\includegraphics[scale=.2]{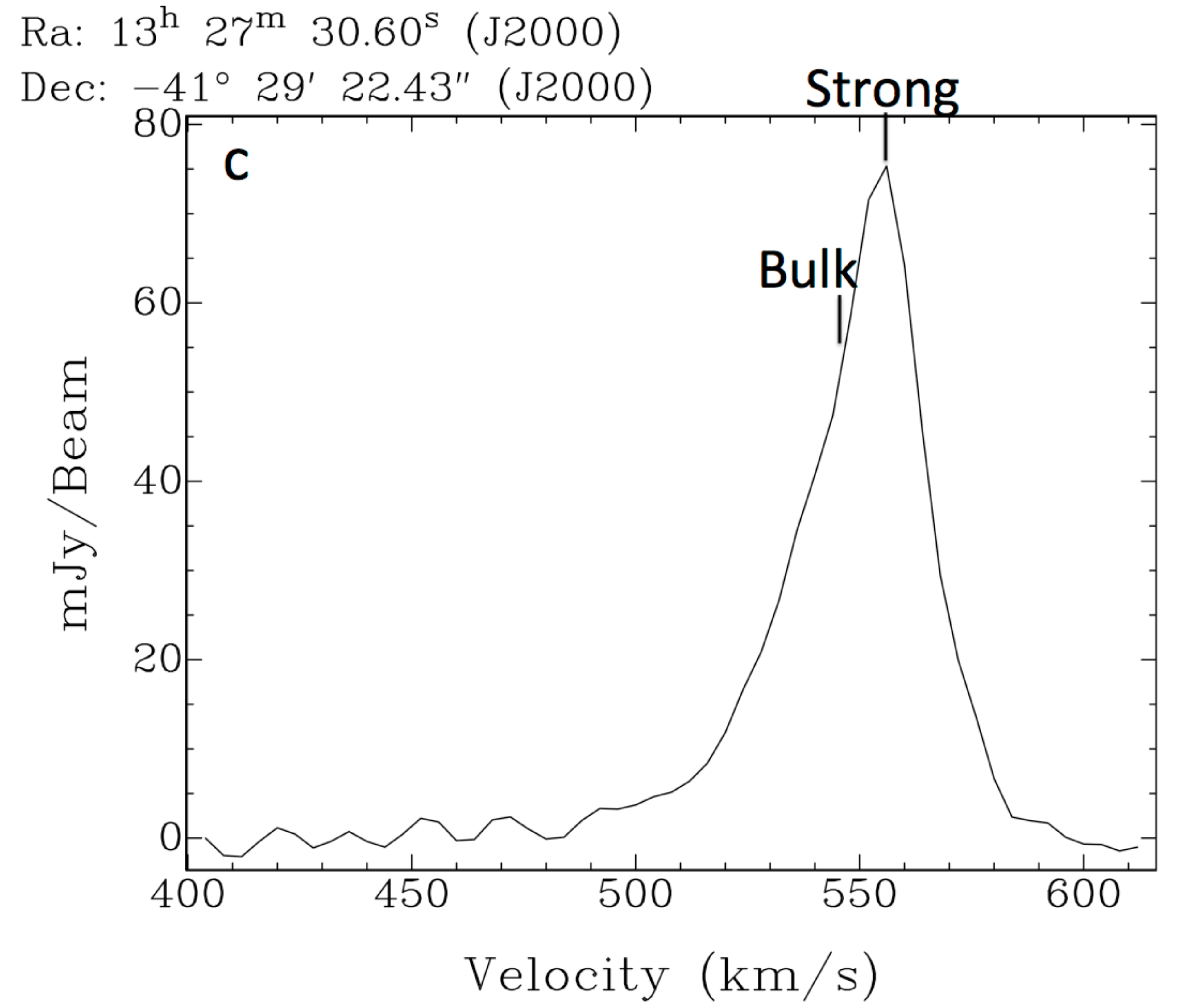}\label{fig:7c}
\hfil
\includegraphics[scale=.2]{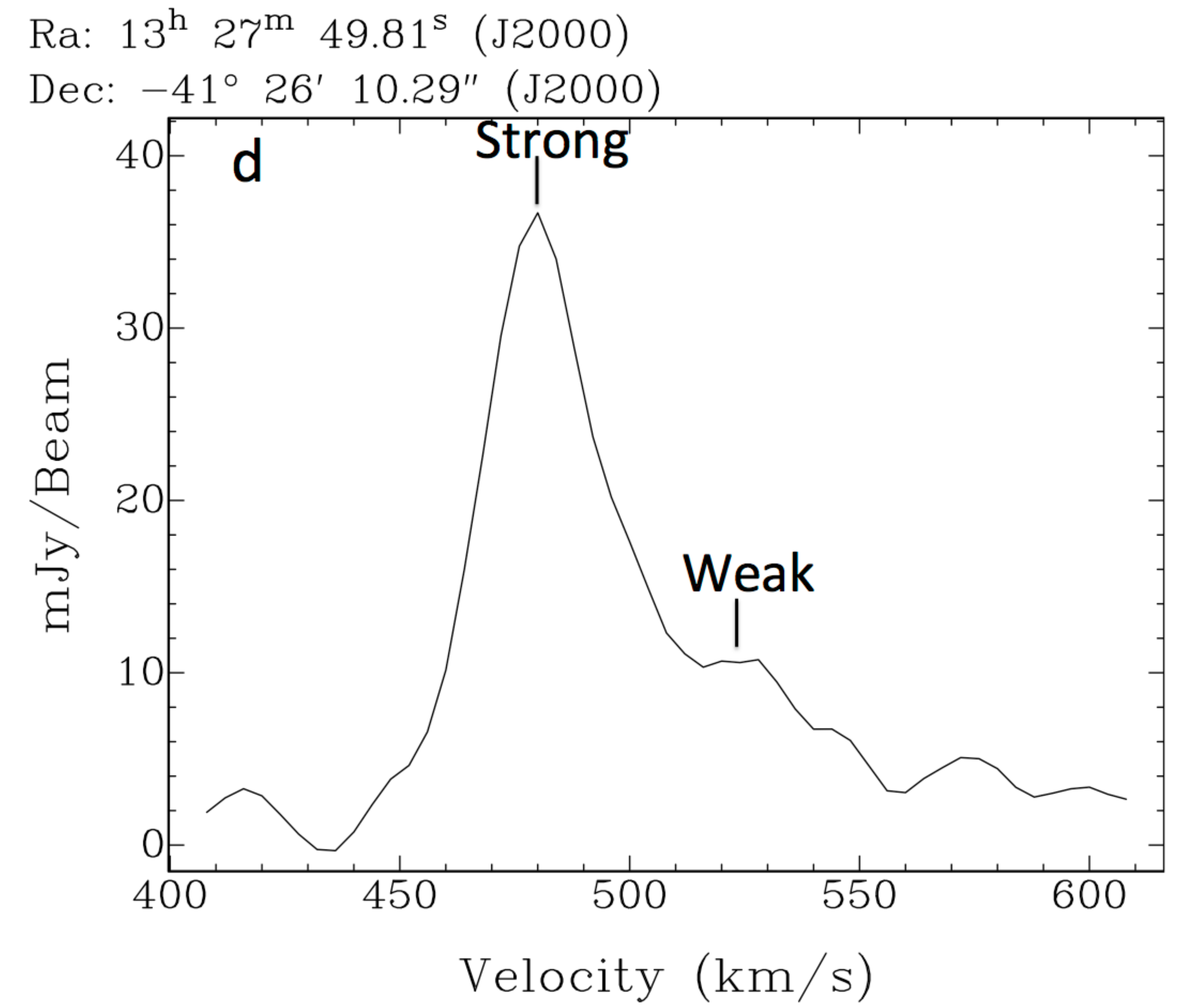}\label{fig:7d}
\caption{Line profiles extracted from the \hi\ data cube to demonstrate the separation of bulk from non-circular motions.  The letters in the upper left corner of each panel correspond to the locations where the line profiles were extracted and are shown in Figure \ref{fig:5}.
The profile in panel (a) shows an example of a single Gaussian as the best fit model, thus, only the bulk velocity field in Figure \ref{fig:5a} contains emission at this location. Panel (b) shows an example of two Gaussian components with strong non-circular motion gas as the higher intensity peak, while panel (c) shows an example of weak non-circular motion gas as the lower intensity peak compared to the bulk motion velocity. Panel (d) shows an example of both strong and weak non-circular velocity gas with no bulk motion in the tail. }
\label{fig:7}
\end{figure*}

 In order to separate bulk from non-circular motions, we fit one or two Gaussian profiles to each \hi\ line profile at every spatial pixel in the data cube. The double Gaussian decomposition code uses a Bayesian inference criterion (BIC) to determine whether a single or double Gaussian profile is best fit to each \hi\ line profile. If two Gaussians is preferred, then the velocities obtained from the two peaks are compared to the model velocity field (Figure \ref{fig:5d}).  The velocity that most closely matches the velocity in the model at the corresponding spatial location is placed into the bulk velocity field (Figure \ref{fig:5a}) while the velocity that deviates more from the model is placed into one of two non-circular motion maps.  The \emph{strong} non-circular motion map (Figure \ref{fig:5b}), contains all  velocities which have intensity peaks \emph{greater} than the bulk velocity peak intensity (see Figure \ref{fig:7c}).  The \emph{weak} non-circular motion map (Figure \ref{fig:5c}), contains all outlying velocities which have intensity peaks \emph{lower} than the bulk velocity peak intensity (see Figure \ref{fig:7b}).  Interestingly, there is a swath of \hi\ in the tail that is blank in the bulk velocity field, yet contains both strong and weak non-circular motions (see Figure \ref{fig:7d}). 
 Perhaps this non-circular motion \hi\ gas is the result of stripped material from either ram pressure or tidal forces.  \color{black}

%
%

			\subsubsection{Tilted-ring model and \hi\ rotation curve}

\begin{figure*}
\includegraphics[scale=.35]{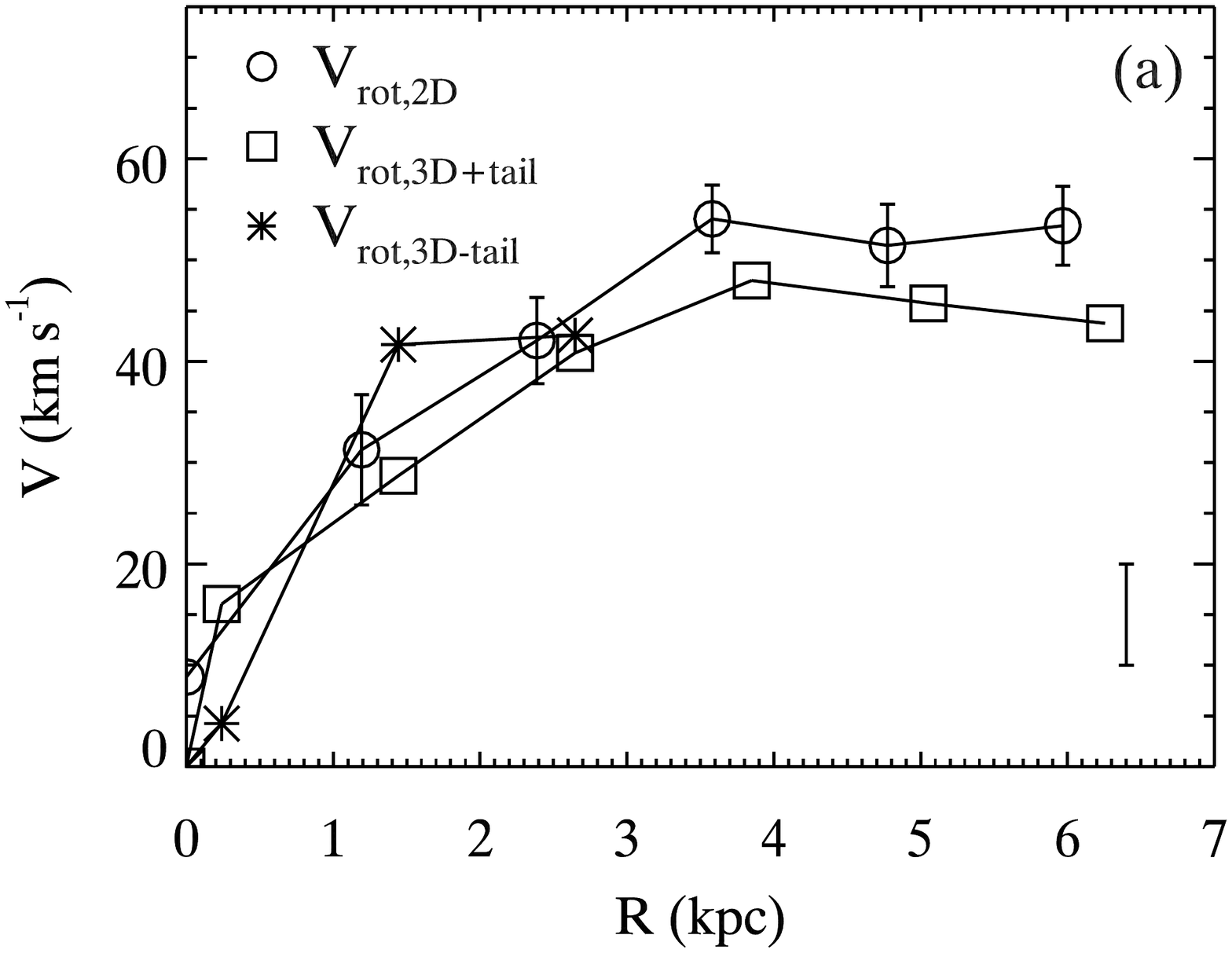}
\hfil
\includegraphics[scale=.35]{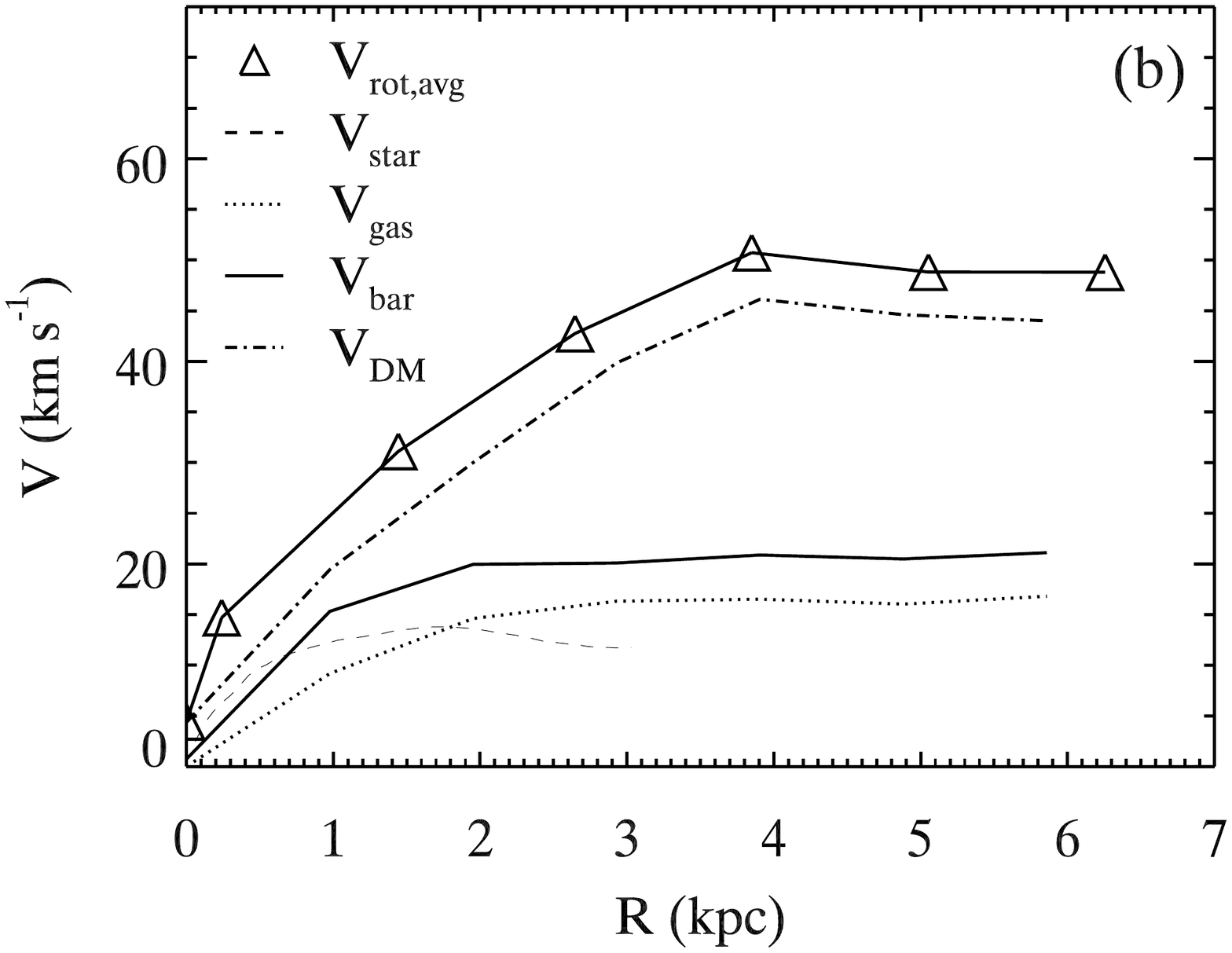}
\caption{(a): Rotation curves for ESO 324-G024 as determined from applying the tilted-ring model to the two-dimensional bulk velocity field shown in Figure \ref{fig:5} ($V_{\rm rot,2D}$, open circles) and from the three-dimensional tilted-ring model, discussed in Section \ref{3Dmod}, for the disk model with the \hi\ tail ($V_{\rm rot,3D+tail}$, open squares) and disk model without the \hi\ tail ($V_{\rm rot,3D-tail}$, asterisks).  (b): The velocity contributions to the average rotation curve ($V_{\rm rot,avg}$, triangles) derived from modeling the mass surface densities of the stars (dashed line) and gas (dotted line) together with the total baryonic contribution determined from the combined stars plus gas velocities (solid line).  The dot-dashed line shows the dark matter contribution to $V_{\rm rot,avg}$.}
\label{fig:8}
\end{figure*}

		
We use the bulk velocity field from Figure \ref{fig:5} and apply a tilted-ring model to derive the rotation curve of ESO 324-G024. We blank any residual noise in the bulk velocity field by using the integrated intensity map from Figure \ref{fig:2} as a mask. Then, we use the GIPSY\footnote{The Groningen Image Processing SYstem (GIPSY) has been developed by the Kapteyn Astronomical Institute.} task {\sc rotcur} to fit consecutive rings with spatial thicknesses of 66$\arcsec$, which is equal to the resolution.  
The tilted-ring model solves a set of orbital parameters for each ring independently initially and these parameters are XPOS, YPOS (the galaxy kinematic centre position), INCL (the inclination of the disk), PA (kinematic major axis of rotation), VEXP (expansion velocity, which we fix to zero), VSYS (the systemic velocity), and VROT (the rotation of the disk).  Each parameter is fixed one at a time until all that remains is VROT.  This is our initial estimate of the rotation curve and its corresponding kinematic parameters.  Once achieved, we then allow one additional kinematic parameter to vary at a time and cycle through the tilted ring model with multiple iterations in order to hone in on a convergent solution for each parameter.  The final values for the orbital elements are given in Table \ref{tab:2} and the resulting rotation curve is shown in Figure \ref{fig:8}.  We calculate the errors on the two-dimensional rotation curve using the following equation from \citet{deb08}:
\begin{equation}
{\rm error}_{V{\rm rot}} = \left(\frac{(V_{\rm rot, approaching} - V_{\rm rot, receding})^2}{16.0} + V_{\rm rot, \sigma}^2\right)^{1/2}
\end{equation}
where $V_{\rm rot, approaching}$ is the rotation velocity of the approaching side, $V_{\rm rot, receding}$ is the rotation velocity of the receding side, and $V_{\rm rot, \sigma}$ is the fitted error on the rotation velocity derived using both sides.
For more details on the two-dimensional tilted-ring model, see e.g., \citet{rog74, beg87, beg89, deb08, oh08, oh11, oh14}.

\begin{table*}
\centering
\caption{\hi\ kinematic parameters for ESO 324-G024}
  \begin{tabular}{lcc}
  \hline
\large  Parameter & \large 2-D Model&\large 3-D+tail Model\\
   \hline
Kinematic centre ($X_{\rm pos}$, $Y_{\rm pos}$) & (13$^{\rm h}$27$^{\rm m}$35.9$^{\rm s}$, $-$41$\arcdeg$28$\arcmin$42.0$\arcsec$)&(13$^{\rm h}$27$^{\rm m}$37.9$^{\rm s}$, $-$41$\arcdeg$29$\arcmin$3.0$\arcsec$)\\
Kinematic Major Axis PA & 253$\arcdeg$ &238$\arcdeg$\\
Inclination, $i_{\rm HI}$ &58$\arcdeg$ & 64$\arcdeg$\\
$V_{\rm sys}$ (\kms) &527& 520 \\
$V_{\rm max}$ (\kms) &54 & 48\\
$V_{\rm max}$sin($i_{\rm HI}$) & 46 & 43\\
$R_{\rm max}$ (kpc) &5.97 &6.26\\
$M_{\rm dyn}$ (M$_{\sun}$) & 4.0 $\times$ 10$^9$ & 3.4 $\times$ 10$^9$\\
\hline
\label{tab:2}
\end{tabular}
\end{table*}

	\subsection{Three-dimensional kinematic model}\label{3Dmod}
	
In order to get a better understanding of the kinematics in ESO 324-G024 we also construct a tilted ring model directly from the data cube, i.e. in three dimensions. For this we use the Tilted Ring Fitting Code 
\citep[T{\sc i}R{\sc i}F{\sc i}C,][]{Jozsa2007}. \color{black} Because there are only a few independent resolution elements over the morphological major axis of the galaxy, we limit the amount of parameters to be fit. The typical fitting process is described in \citet{Kamphuis2011} and \citet{Zschaechner2011}. For ESO 324-G024, we let the surface brightness (SBR), rotational velocities (VROT), position angle (PA) and inclination (INCL) vary as functions of radius and additionally fit the scale height (Z0) as a single value.

\begin{figure*}
   \centering
   \includegraphics[width=6 cm ]{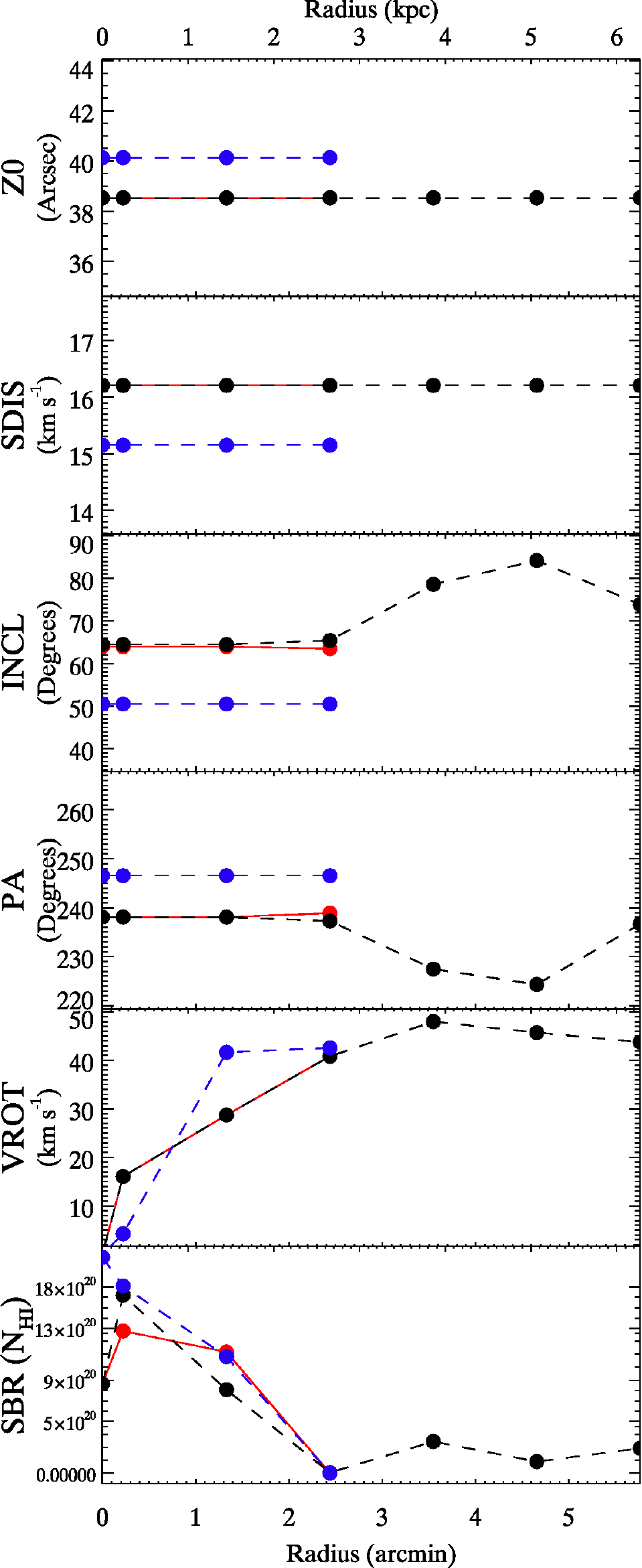} 
   \hspace*{0.5 cm}
    \includegraphics[width=6.2 cm]{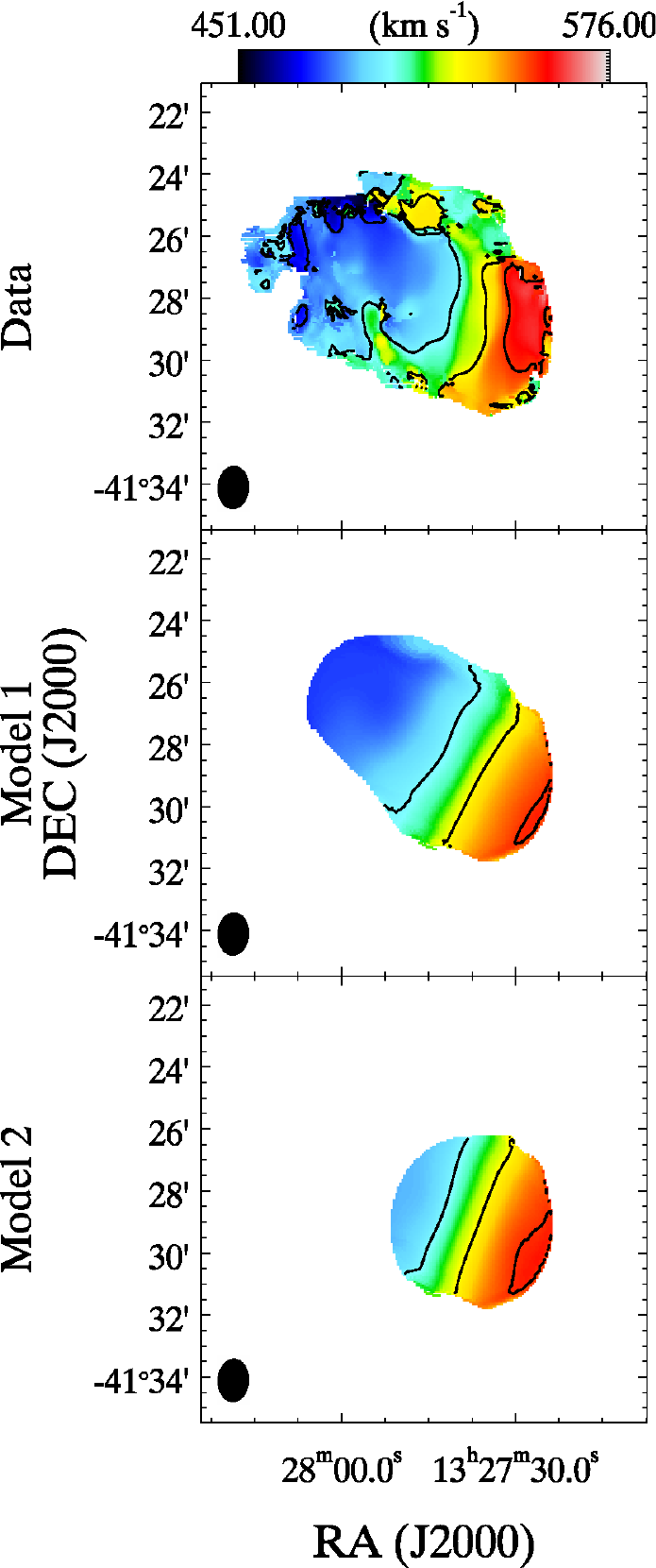} 
   \caption{Left Panels: Parameters of the tilted ring model as a function of radius; the red lines are the receding (western) side and the black-dashed lines are the approaching (eastern) side in the model where we fit the receding and approaching sides independently and included the emission from the \hi\ tail (Model 1).  
   For several of the parameters, the red and black-dashed lines overlap, thus indicating agreement between the receding and approaching sides. \color{black} The blue lines represent the model where we fit the disk as a whole without the emission from \hi\ tail (Model 2). Right Panels: \hi\ velocity fields for the data (top), the three dimensional model that includes the \hi\ tail (middle), and the three dimensional model without the \hi\ tail (bottom). }
   \label{Par+vel}
\end{figure*}

We initially assume that all the \hi\ emission is on circular orbits and therefore, we fit all the orbital parameters for the approaching and receding sides of the galaxy independently with the exception of the rotation curve, which is kept the same for both sides. Figure \ref{Par+vel} shows the final parameters of this model in the panels on the left and on the right, the velocity fields for the data and the models, as derived with a Gaussian-Hermite fit. 

The model where the disk is fit as a whole and includes the tail (Model 1, Figure \ref{Par+vel}) performs well. The centre of the model at $V_{\rm sys}$=519.8 \kms, RA=13$^{\rm h}$ 27$^{\rm m}$ 37.9$^{\rm s}$ and DEC=$-$41\arcdeg\ 29\arcmin\ 3\arcsec\ is similar to the optical position centre. This is remarkable as we are treating the tail as being part of the circular rotation in this galaxy. Therefore, one would expect the tail to drive the model centre away from the optical centre if the velocities were very distinct from circular. This model gives a reasonable fit up to a radius of 2$\farcm5$ and beyond this, the PA and inclination angles begin to vary, which may suggest that this model does not match the data well at these radii (see Figure \ref{Par+vel}).

We conclude that the gas in the tail is likely stripped from the galaxy because the velocities in the tail follow those expected from circular motion with a rotational centre position that is not distinct from the optical center. 

As the tilted ring model should only be applied to emission on circular orbits we refit the data with a single circular disk which is truncated at 2.5\arcmin. In this way we try to fit the main disk and ignore the tail. In order to not artificially shift the model towards the tail we keep the positional centre (RA, DEC) fixed to the position found in the previous model and only refit it, by itself, after finding a good fit to the disk. Additionally, we only fit a flat disk (PA, INCL one value) as there are only a few resolution elements over the morphological major axis of this disk.

Figure \ref{Par+vel} shows the model velocity field in the bottom right panel (Model 2) and the parameters are shown by the blue dashed lines in the left hand panels. The first thing that becomes clear is that the PA appears offset when compared to the velocity field of the data (top left panel). However, this seems to be an artifact of the representation in velocity fields as a visual inspection shows that the emission in the data is traced well by this model. The rotation curve is not corrected for asymmetric drift because this correction results in roughly a channel width for ESO 324-G024, i.e. $\sim$4  km s$^{-1}$. One interesting, albeit poorly constrained, parameter in the model is the scale height, which is nearly 40\arcsec\ (720 pc) meaning the disk appears very puffed up.

\begin{figure*}
   \centering
   \includegraphics[scale=0.5]{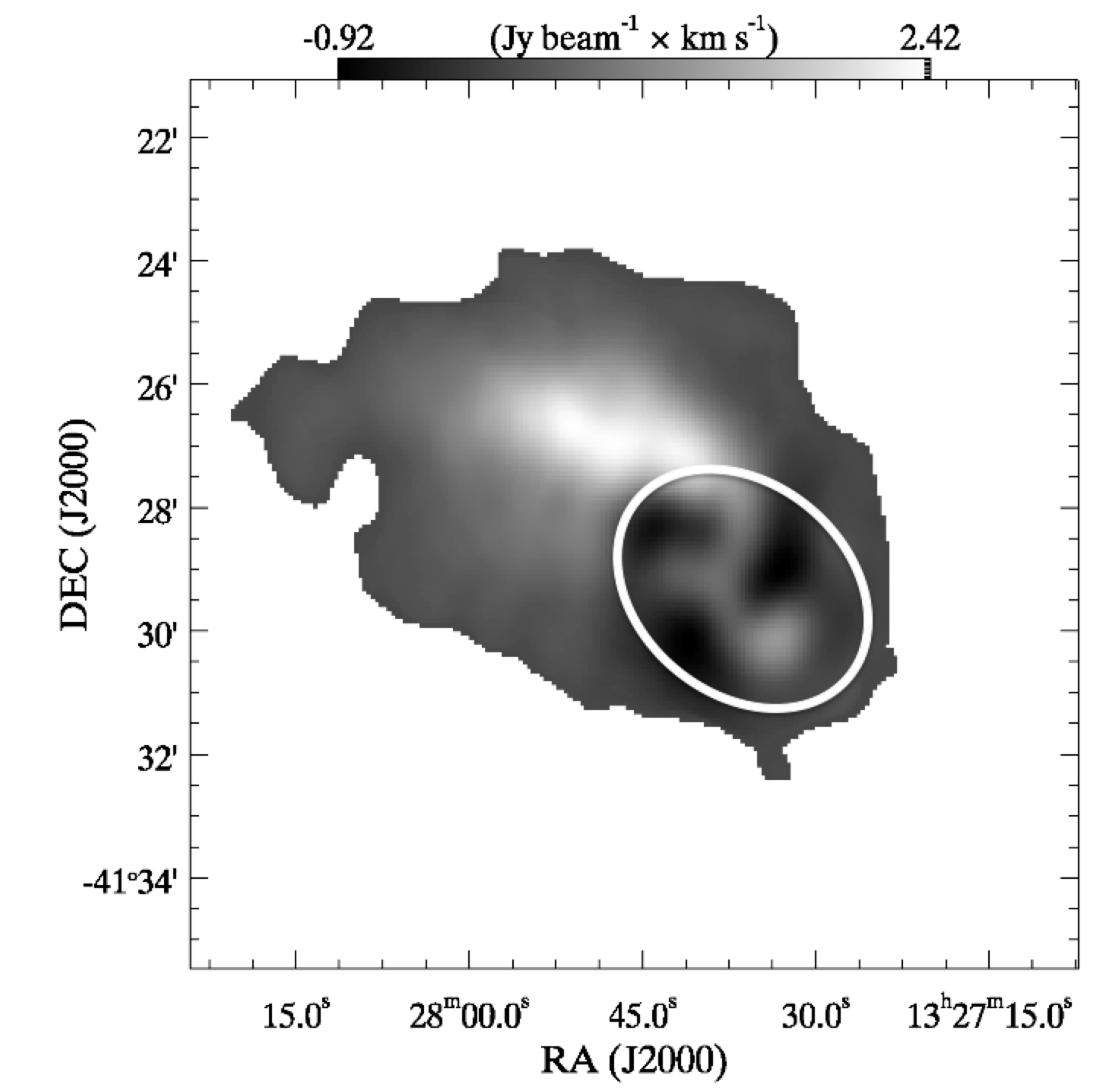} 
   \caption{Integrated intensity map of the data minus the one disk model. 
   The \hi\ emission is predominantly from the tail in ESO 324-G024 and the white oval demarcates the optical $B-$band radius at 25 mag $\rm arcsec^{-2}$.\color{black}}
   \label{residual}
\end{figure*}

By subtracting Model 2 (bottom 
right \color{black}panel in Figure \ref{Par+vel}) from the data (top 
right \color{black}panel in Figure \ref{Par+vel}) we can make some estimates on the mass of tail. If we apply the data mask as created by SoFiA \citep{Serra2015} to the residual cube, we are left with the emission shown in Figure \ref{residual}. From this process, we estimate that there is a total of 7.4 $\times\ 10^{7}$ \msun\ of \hi\ gas in the tail, which is $\sim$36\%\ of the total emission.  It must be noted that this is a very crude approximation. Additionally, from the surface brightness in Model 1 in Figure \ref{Par+vel}, we get an average surface density $\Sigma_g$ of 8.8 M$_{\odot}$ pc$^{-2}$ for the inner disk and 1.7 M$_{\odot}$ pc$^{-2}$  in the tail for the remainder of the disk from the two disk model.

In conclusion, we find that the kinematics of the tail in ESO 324-G024 closely resembles the velocities of those expected from circular rotation in the disk.  Therefore, we find that the ram pressure is the most likely mechanism responsible for the creation of this \hi\ feature.

	\subsection{Comparing the two- and three-dimensional kinematic models}

As we see in Figure \ref{fig:8}, the rotation curves derived from the two- and three-dimensional models agree well to within the errors.
Upon comparing the kinematic parameters of the models (see Table \ref{tab:2}), we find that the two-dimensional model finds a higher PA and a lower $i$ than the three-dimensional model, yet the \vmax sin($i$) values agree to within a channel width.

The differences between the two- and three-dimensional models are most likely attributed to the low surface brightness in the tail and the poor spatial resolution of the \hi\ data in the disk of ESO 324-G024. 
According to \citet{bos78}, the minimum number of spatial resolution elements within the Holmberg radius (i.e., the $B-$band optical radius at a surface brightness of 26.5 mag arcsec$^{-2}$) required for determining reliable two-dimensional rotation curves is approximately seven. Kamphuis et al.\ 
(in preparation) finds approximately six resolution elements are required for reliable three-dimensional modeling.  ESO 324-G024, on the other hand, contains only about four spatial resolution elements over its Holmberg radius.

It is not clear if the tail in ESO 324-G024 is, in fact, rotating in circular motion as the three-dimensional model suggests
 and therefore, it may not be appropriate to model this gas with the tilted-ring model.  The two-dimensional model finds that a large portion of the \hi\ tail  contains both strong and weak non-circular motions and no bulk motion.  The three-dimensional model, on the other hand, fits for the surface brightness in addition to the orbital parameters.  In the region of the \hi\ tail, we find that the \hi\ is at low surface brightness with velocities over a wide range of channels.  Figure \ref{fig:7d} shows a line profile with a peak at only 40 mJy at the location of the tail and has a total velocity range of roughly 100 \kms.  Thus, when modeling this low surface brightness, broad \hi\ gas in three-dimensions, the model finds \hi\ that agrees with circular velocity at the locations of the tail.  Therefore, it may appear as though the two- and three-dimensional models do not agree in the tail, but, this is simply due to the broad, low surface brightness \hi\ at that location. The poor spatial resolution together with the broad \hi\ tail emission means that the shape of the rotation curves in Figure \ref{fig:8} may not be reliable, however, our maximum rotation velocity is robust as the two models agree to within the channel resolution.  
 
\begin{figure*}
\centering
\subfigure[PA = 254$\arcdeg$\label{fig:PVb}]
{\includegraphics[scale=.95]{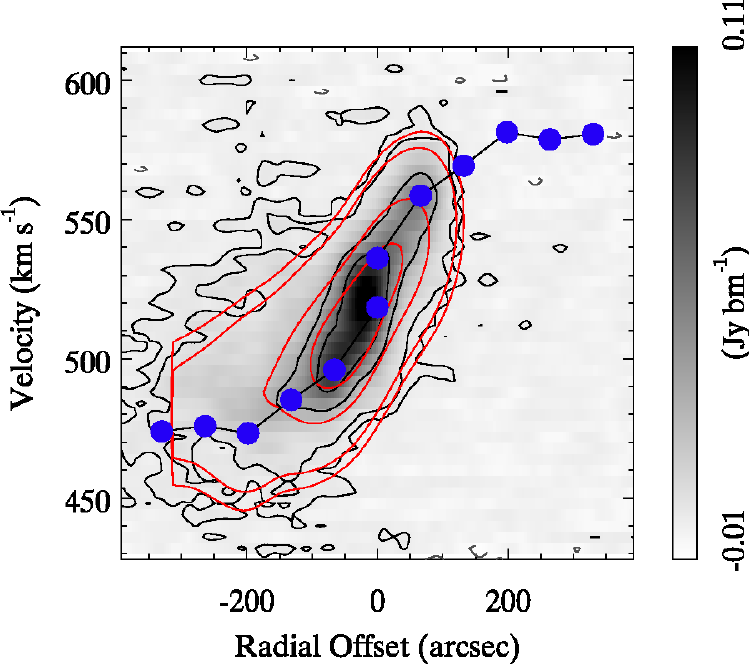}}
\subfigure[PA = 238$\arcdeg$ \label{fig:PVa}]{\includegraphics[scale=.95]{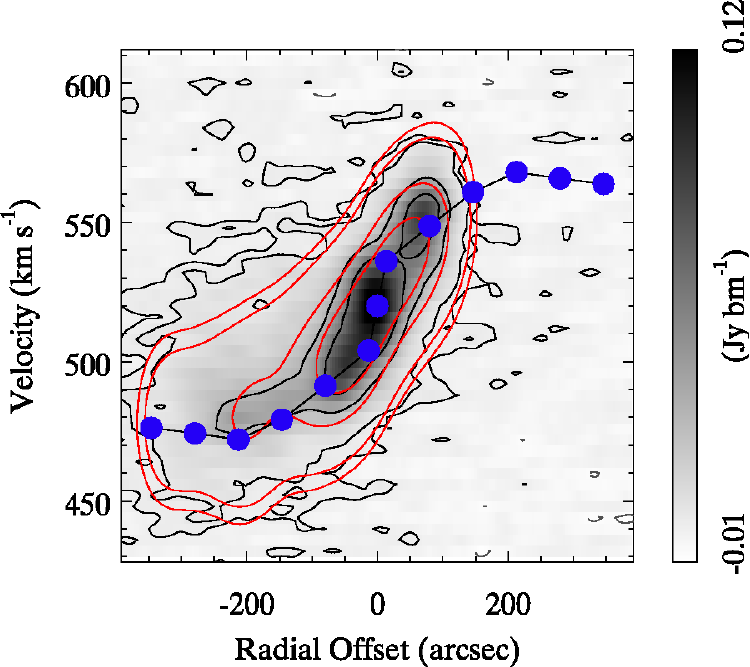}}
\subfigure[PA = 336$\arcdeg$]
{\includegraphics[scale=.95]{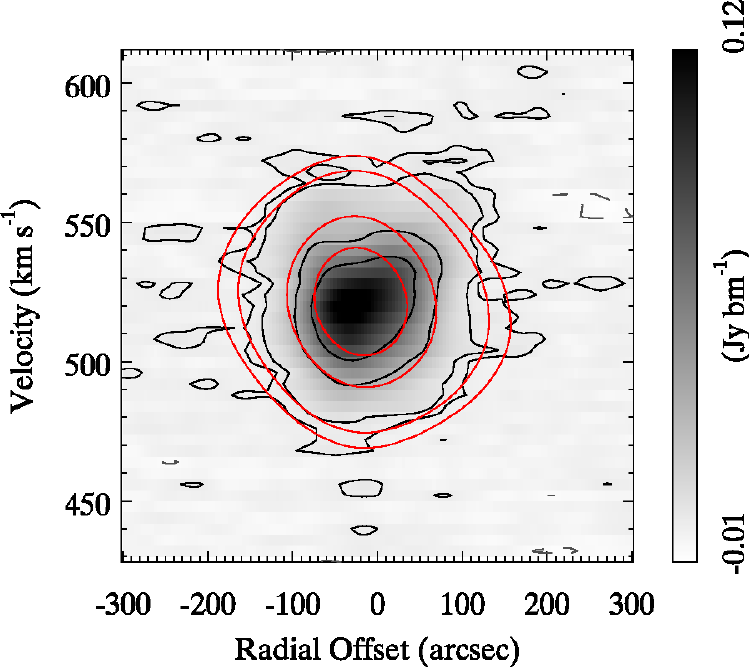}\label{fig:PVc}}
\subfigure[PA = 250$\arcdeg$]
{\includegraphics[scale=.95]{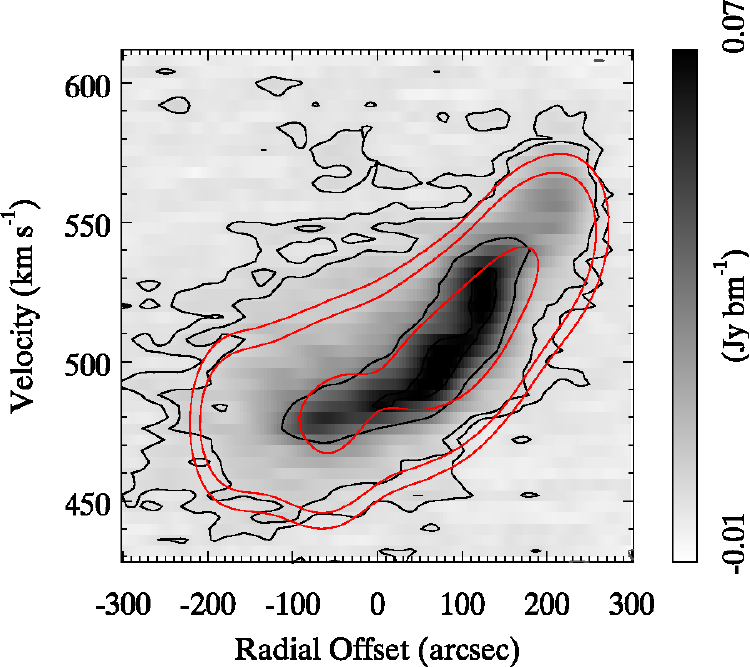}\label{fig:PVd}}
\caption{Position-velocity (P-V) diagrams extracted along four position angles (PAs) measured east from north with a width of 66$\arcsec$ equal to the spatial resolution of the  \hi\ data. Panel (a) is the kinematic major axis of the two-dimensional model, (b) is the kinematic major axis of the three-dimensional model, (c) is the kinematic minor axis of the average between the two- and three-dimensional models, and (d) is the region of the \hi\ tail.  The center positions of the diagrams in panels (a) and (b) are the kinematic centers shown in Table \ref{tab:2}.  The center position for panel (c) is the average of the kinematic center positions derived from the two- and three-dimensional models.  The center position of panel (d) is at (13:27:47.7, $-$41:27:10.3), which is to show the emission in the \hi\ tail. In all panels, the black contours show the \hi\ intensities while the red contours show the three-dimensional kinematic model intensities (at the same intensity levels) for the model that includes the \hi\ tail.}
\label{fig:PV}
\end{figure*}

 Figure \ref{fig:PV} shows four position-velocity (P-V) diagrams extracted from the \hi\ data cube.  In panels (a) and (b), we overlay the respective rotation curves for the two- and three-dimensional models, respectively.  What is evident in these P-V diagrams is that along the position angle fitted in the 3D (model 1) more of the gas in the the tail is captured. Indicating that this model is the model that is most affected by the gas in the tail. Both models show no emission on the receding side of the galaxy beyond a radius of $\sim$150$\arcsec$ (2.7 kpc), which implies that the \hi\ tail is responsible for determining the circular velocities at these large radii.
  
If we accept that the tail is rotating in a circular motion and thus, is appropriately fitted with the tilted-ring analysis, then we estimate a maximum rotational velocity for ESO 324-G024 by averaging the velocities of the two- and three-dimensional models together to produce the rotation curve shown in panel (b) of Figure \ref{fig:8} and find $<V_{\rm max}>$ $\approx$ 51 $\pm$ 5 \kms.  
The error on this value and the error bar shown in the bottom right corner of panel (a) in Figure \ref{fig:8} is representative of the difference between the inclinations of the two- and three-dimensional models when applied to $<V_{\rm max}>$ sin($i_{HI}$).  
From the virial theorem, we find a total dynamical mass of 3.7 $\times$ 10$^{9}$ \msun\ at a radius of 6.1 kpc, the average radius at which $<V_{\rm max}>$ was determined.




\section{Discussion}\label{sec:discuss}

In the previous sections, we have determined that ESO 324-G024 must be behind the northern radio lobe of Cen A based on the lack of depolarized signal in the polarized intensity and rotation measure maps of Cen A at the location of ESO 324-G024.   We have also studied the kinematics of the \hi\ in ESO 324-G024 and find that the galaxy has a non-circular motion \hi\ cloud toward the end of a long, 3.5 kpc tail extending away from the core of NGC 5128.   The \hi\ tail appears to have motions suggesting that it is rotating circularly similar to the rest of the disk.  Here, we explore the implications of these results.

	\subsection{The effect of tidal forces on ESO 324-G024}\label{sec:gravity}
	
To better understand what the role of the gravitational forces are on ESO 324-G024, we begin by looking at the possibilities of harassment by other galaxies.  
\citet{smi10} study the affects of harassment on Virgo Cluster dIrr galaxies and define harassment from the literature as ``the effects of repeated and numerous long range tidal encounters in the potential well of the cluster.'' In this sense, as \color{black} ESO 324-G024 is in a small galaxy group, harassment events are unlikely because of the low frequency of encounters and the low
velocities at which these encounters occur; the average velocity
dispersion in the Centaurus A galaxy group is only 114 \kms\ \citep{van00}  as determined
from the radial velocities of 30 group members.  This is much lower than the typical velocities in a cluster,  e.g. 700 \kms\ in the Virgo Cluster \citep{bin93}, and therefore we rule out harassment as a potential cause of the \hi\ tail in ESO 324-G024.
	

In most popular models and observations of tidal interactions in gas-rich galaxy mergers, tidal interaction creates two tidal tails, approximately symmetric, at the near and far end of the galaxy towards the other galaxies' direction, respectively \citep[e.g.,][]{hib99, hib00a, hib00b}.  This is not the case in ESO 324-G024, which has only a single \hi\ tail, making this feature likely not from tidal forces as no symmetric counter tail is observed.  

According to the models by \citet{smi10} for the Virgo Cluster dIrrs, the effects of tidal forces produce only mild signatures in the stars, which include short-lived tidal tails and very low surface brightness stellar streams. They also show that tidal encounters can induce \hi\  tidal structures in the gaseous disks and bar-like formations in the central regions.  During these episodes, the models show enhancements in SFRs.  The authors state that these SFR increases can be periodic as the tidal encounters stretch out and lower the gas densities thus, lower SFRs, and then compress the gas, which in turn increases SFRs.  However, these features seem to be short-lived.  

For ESO 324-G024, we have previously shown that its SFR is in the realm of a weak starburst dIrr, thus, tidal forces may be a potential mechanism for this episode of star formation.  We notice a slight `S'-like pattern in the \hi\ intensity contours (see Figure \ref{fig:2}), which may be an indication of weak spiral arm-like structures in the gas disk.  However, the spatial resolution of the \hi\ data is not enough to definitely determine if this `S'-like  morphology is real.
On the other hand, we see no clear evidence for tidal tails in the stellar disk or of a bar-like formation in the central regions in either the \hi\ or stars.



An analytical approach to determining whether tidal forces are responsible for creating the \hi\ tail is to estimate the tidal radius, $r_{\rm T}$, of ESO 324-G024.  According to \citet{rea06}, the tidal radius is the radius at which a star belonging to a satellite system orbiting a host galaxy becomes unbound or stripped from the satellite and instead becomes bound to the host. We use the Jacobi approximation that assumes that the interacting systems are point masses such that $M >> m$ and determine the tidal radius in a similar manner to \citet{rea06} and \citet[][see \S\,7.3]{bin87}:
\begin{equation}\label{eq:tid}
r_{\rm T} \approx\ R\left(\frac{m}{M(3+e)}\right)^{1/3}
\end{equation}
where $R$ = 125 kpc is the true distance from ESO 324-G024 to the centre of NGC 5128 using a projected distance of 104 kpc and a distance along the line of sight of 70 kpc  (see Figure \ref{fig:4}), $m$ = 3.7 $\times$ 10$^{9}$ \msun\ is the mass of ESO 324-G024, $M$ = 1.3 $\times$ 10$^{12}$ \msun\ is the mass of NGC 5128 \citep{woo07} and $e$ is the eccentricity of the orbit of ESO 324-G024 about NGC 5128.  If we assume that both systems are in circular orbits about the center of mass (i.e., $e$ = 0), then we find $r_{\rm T}$ = 12 kpc, twice the maximum radius of ESO 324-G024, which is at most 6 kpc measured from the centre to the end of the \hi\ tail. If we assume that ESO 324-G024 is on a hyperbolic orbit about NGC 5128 where $e$ = 1, we find that $r_{\rm T}$ = 11 kpc, which is still beyond the maximum \hi\ radius.   Similarly, if we assume the host galaxy is a singular isothermal sphere rather than a point mass, then $r_{\rm T}$ increases by a factor of 1.145 \citep{bin87}.

In addition, we determined the relaxation timescale, which is the time necessary for a galaxy to have its orbit perturbed by a tidal interaction and is derived with the following from \citet{bos06}:
\begin{equation}\label{relax}
t_{\rm relax} = 0.1\left(\frac{D}{\delta V}\right)\left(\frac{N_{\rm gal}}{{\rm ln}\ N_{\rm gal}}\right)
\end{equation}
where $D$ = 526 kpc \citep{kar02} is the diameter of the group, $\delta V$ = 114 \kms\ \citep{van00} is the group velocity dispersion, and $N_{\rm gal}$ = 31 \citep{van00, kar02} is the number of member galaxies in the group. We find $t_{\rm relax}$ = 4 Gyr for ESO 324-G024, which is larger than the crossing time of 3 Gyr \citep{kar02} for the Centaurus A galaxy group and indicates that tidal encounters are likely not common for ESO 324-G024.
\color{black}

It has been suggested in the literature that one of the likely mechanisms for creating starbursts in dwarfs is the tidal interaction of a dwarf-dwarf galaxy pair \citep[][and references therein]{mcq10a, mcq10b, joh12, joh13, nid13}.   The closest dwarf galaxy to ESO 324-G024 is the dwarf spheroidal galaxy KK197 \citep{kar13}, which has a distance of 3.87 Mpc as determined from the tip of the red giant branch.  
KK197 is at a projected distance of 114 kpc from ESO 324-G024 and if we assume that these dwarfs are moving at the mean velocity dispersion of the Centaurus A galaxy group, 114 \kms\ \citep{van00}, then it's been at least 1 Gyr since the two were coincident.  This makes the likelihood of an interaction with KK197 an improbable situation to have created an \hi\ tail because gaseous tidal tails generally exist for $<$ 1 Gyr from the initial galaxy encounter before they dissipate into the intergalactic medium \citep{too77}.

In conclusion, tidal forces alone do not appear to be enough for the creation of the \hi\ tail in ESO 324-G024.


	\subsection{The effect of ram pressure on ESO 324-G024}\label{sec:rampres}
It was first suggested by \citet{fri97} that ESO 324-G024 is being affected by ram pressure and undergoing stripping of its \hi\ gas based on its striking morphology.  We explore this situation further and expand on their work in this section.  

To begin, we derive the peculiar velocity, $V_{\rm pec}$, of ESO 324-G024 by first determining the expected velocity from the Hubble flow, $V_{\rm Hub}$, using a Hubble constant of 69.32 $\pm$ 0.80 \kms Mpc$^{-1}$ \citep{ben13} and find $V_{\rm Hub}$ = 260 \kms. This velocity subtracted from the Local Group standard of rest velocity, $V_{\rm LG}$ = 282 \kms\ (ascertained from the average of the systemic velocities given in Table \ref{tab:2} and transformed in NED\footnote{NASA/IPAC Extragalactic Database}), implies that ESO 324-G024 has $V_{\rm pec}$ = $V_{\rm LG} - V_{\rm Hub}$ = 22 \kms. This slightly positive $V_{\rm pec}$ 
together with the head-tail morphology suggests that ESO 324-G024 is traveling through the galaxy group in the direction of the head away from our line of sight and thus, the 
orientation of ESO 324-G024 is such that the tail is closest to us while the head is furthest from us.  Because the dwarf is behind the northern radio lobe of Cen A, it is plausible that the dwarf passed through the lobe and is now on the far side.  However, depending on the lobe's inclination to our line of sight and the orbital path of ESO 324-G024, it may have only passed through part of the lobe or grazed its posterior outer edge.  

In support of ram pressure as the dominant force acting on ESO 324-G024 and creating the \hi\ tail, we examine the individual channel maps shown in Figure \ref{fig:chan}. What is astonishing in this figure is that at the location of the western edge of the optical disk (delineated by the black oval), the \hi\ intensity is sharply truncated at all velocities.  We can also see evidence for \hi\ compression as the black intensity contours in the channel maps are packed closely together near the western edge of the optical disk and are more separated toward the tail.  
These features imply that an IGM wind is likely blowing in the face of ESO 324-G024 as it moves through the Cen A galaxy group.

The minimum amount of ram pressure that ESO 324-G024 would need to cause stripping can be estimated from \citet[][equations 61 \& 62]{gun72}:
\begin{equation}\label{eq:ram}
p_{\rm ram}^{min} = \rho_{\rm IGM}\ v_{\rm gal}^2 > 2 \pi G \Sigma_{gas} \Sigma_* = 10^{-3}\ \rm{cm}^{-2}
\end{equation}
where $\rho_{\rm IGM}$ is the density of the intergalactic medium (IGM), $v_{\rm gal}$ = 197 \kms\ is the velocity dispersion of the group \citep[114 \kms;][]{van00} times the square root of three to account for the motion of the galaxy in the plane of the sky, $G$ is the gravitational constant, $\Sigma_{gas}$ = 4.9 $\times$ 10$^{20}$ cm$^{-2}$ is the average gas mass surface density over the disk, and $\Sigma_*$ = 9.3 \msun\ pc$^{-2}$ is the average stellar mass surface density out to the optical R$_{\rm 25}$ radius determined from the profile in Figure \ref{surfden}.

Comparing this IGM density with the thermal gas number density inside the radio lobe of Cen A of $\sim$10$^{-4}$ cm$^{-3}$ from \citet{osu13}, it would appear that the medium inside the lobe is not dense enough to strip the \hi\ gas of ESO 324-G024. It is important to note, however, that the thermal gas number density derived by \citet{osu13} is a \emph {volume average} and that the density at the edge of the lobe could be higher.  \citet{kra09} determine a density of $\sim10^{-3}$ cm$^{-3}$ at a radius of roughly 35 kpc from the core of NGC 5128. \citet{bou07} study the \hi\ content in some of the companion galaxies in the Centaurus A galaxy group and find that there is a lack of \hi\ detected in their targets.  They speculate that the galaxies without \hi\ were stripped of their gaseous components and they determine that an IGM density of $\sim10^{-3}$ cm$^{-3}$ is necessary to cause such stripping in the galaxies around NGC 5128.  However, \citet{eil14} state that a density of 10$^{-4}$ cm$^{-3}$ could also explain the stripping of the \hi\ content in the companion galaxies of NGC 5128 when following the formalism of \citet{gun72}.  Thus, it is possible that the density of the medium of the lobe through which ESO 324-G024 passed is somewhere between 10$^{-4} - 10^{-3}$ cm$^{-3}$.

Our simple calculation in Equation \ref{eq:ram} does not take into account the velocity of the particles inside the lobe of Cen A.  If we set $\rho_{\rm IGM} = 10^{-4}$ cm$^{-3}$ in equation \ref{eq:ram} and replace $v_{\rm gal}$ with $v_{\rm tot}$, where $v_{\rm tot} = v_{\rm gal} + v_{\rm lobe}$ is the total velocity experienced by ESO 324-G024 as it moves through the radio lobe (its velocity through the galaxy group, $v_{\rm gal}$ plus the velocity of the medium inside the lobe of Cen A, $v_{\rm lobe}$), then we determine that the total velocity required 
for the stripping of the \hi\ disk of ESO 324-G024 is $v_{\rm tot}$ = 330 \kms. If $v_{\rm gal}$ = 197 \kms, then $v_{\rm lobe} =$ 133 \kms.  \citet{eil14} provides some estimates for the outflow speeds that range from 480 -- 9800 \kms\ with the most probable model being the low velocity end of this range.  
It is likely that the wind direction of the lobe is not in perfect alignment with the direction of the galaxy through the IGM, thus there may be only a partial velocity component from the lobe that is actively driving the ram pressure force on ESO 324-G024.  Therefore, the velocity range from \citet{eil14} appears to be in good agreement with our approximate estimate for the lobe velocity acting on ESO 324-G024.
 
In comparison, \citet{smi13} model the ram pressure force on tidal dwarf galaxies with no dark matter halos and compare them to dIrr galaxies with dark matter halos.  
 We find a striking similarity to the model for a dIrr galaxy with a dark matter halo undergoing a ram pressure wind speed of 400 \kms, which is similar to the total velocity estimate of $v_{\rm tot}$ = 330 \kms\ required for ESO 324-G024.
 \citet{smi13} show that dIrr galaxies enveloped in dark matter halos that undergo ram pressure forces with a wind speed of 400 \kms\ have relatively undisturbed stellar disks and long \hi\ tails that matches what we observe in ESO 324-G024. 

The timescale, $\tau$, over which a ram pressure stripping event occurred in ESO 324-G024 can be estimated by simply taking the length of the tail (which is 3.5 kpc as measured from Figure \ref{residual}) and dividing it by the velocity difference, $v_{\rm diff}$ = 54 \kms, between the stripped material and the systemic velocity of the galaxy.  Then, if we assume that $v_{\rm diff}$ is purely in the radial direction and that stripping is constant with time, we determine the minimum time since the onset of the stripping is $\tau \approx$ 63 Myr.  It is likely that $v_{\rm diff}$ is not purely radial, in which case $\tau$ would be higher.  

If ESO 324-G024 is traveling through the galaxy group at 114 \kms, then it has traveled about 7 kpc in 63 Myr.  This distance is roughly equal to the length of the white box in Figure \ref{fig:2}a, which is remarkably small compared to the large lobe of Cen A.  
If ESO 324-G024 had passed through the full depth of the northern radio lobe of Cen A, then it would have been exposed to ram pressure forces from the lobe for $>$ 2 Gyr (the length of time required to travel the full 200 kpc dearth of the lobe at 114 \kms).  This length of time would likely have stripped all of the \hi\ from ESO 324-G024.  Therefore, we conclude that ESO 324-G024 is on its first approach and is just now coming in contact with the posterior outer edge of the northern radio lobe of Cen A.  
In addition, our estimated ram pressure timescale of 63 Myr is similar to what is observed in other cases of ram pressure stripping like the Coma Cluster and the Virgo Cluster where ram pressure occurs quickly over short timescales on the order of $\sim$10$^7$ yr \citep{aba99, bos08}. \color{black}

\citet{ste13} compare \emph{ROSAT} All Sky Survey data with their radio continuum data from PAPER \citep[the Donald C.\ Backer Precision Array for Probing the Epoch of Reionization]{par10} and report on a north-south X-ray filament that was first discovered by \citet{arp94}.    
\citet{arp94} analyzed a 10$\arcdeg\ \times\ 10\arcdeg$ region centered on NGC 5128 using the \emph {ROSAT} X-ray emission data
and discovered a diffuse X-ray filament that extends nearly five degrees on the sky in a northeastern direction at a position angle of about 53$\arcdeg$.  Arp also identifies diffuse X-ray emission north of NGC 5128 at a position angle close to zero.  In Figure 2 of Arp (1994), we see that ESO 324-G024 is coincident with a patch of hard X-ray emission north of NGC 5128. It is argued that this hard X-ray emission is associated with hot ionized gas in the outer edges of the giant radio lobes.  If ESO 324-G024 passed through the lobe of NGC 5128, then it would have passed through this hot ionized gas, which is the likely source for the ram pressure.

\section{Summary \& Conclusions}\label{sec:sum}

\subsection{Geometric Orientation of ESO 324-G024 with respect to NGC 5128}

We find that ESO 324-G024 may be undergoing a starburst episode as defined by \citet{mcq10a, mcq10b} and therefore, likely has enough magneto-ionic material with which to cause depolarization if it were in front of or inside the northern radio lobe of Cen A.
From the 1.4 and 5 GHz continuum data of NGC 5128, we find no depolarization signature in the northern radio lobe at the location of ESO 324-G024 and therefore, we conclude that ESO 324-G024 is most likely behind the northern radio lobe of Cen A.
Because ESO 324-G024 is behind the northern lobe of Cen A, we can place limits on the possible orientations of the lobe and the relative distances of ESO 324-G024 and NGC 5128. One possible scenario (assuming that the distance to ESO 324-G024 and NGC 5128 is the same and that the lobe is as deep as it is wide) is that the northern radio lobe of Cen A is inclined by 60$\arcdeg$ toward our line of sight.

\subsection{Kinematic \hi\ results in ESO 324-G024} 
Using a two-dimensional double Gaussian decomposition technique from \citet{oh08, oh11}, we separated bulk from non-circular motions and find a non-circular motion cloud at the end of the \hi\ tail in ESO 324-G024 that is offset by $-20$ \kms\ from the circular velocity of the model velocity field.
We applied a tilted ring model to the bulk velocity field and find a maximum rotation speed of 54 \kms.
We also applied a tilted ring model directly to the \hi\ data cube of ESO 324-G024, i.e., in three dimensions, and derived a maximum rotation speed of 48 \kms.
 The kinematic modeling highlights that the \hi\ tail of ESO 324-G024 appears to follow the expected circular rotation velocity despite its extended morphology.
If we average together the two maximum rotation speeds from the two independent kinematic models, we determine a maximum rotation velocity of 51 $\pm$ 5 \kms\ for ESO 324-G024, which produces a dynamical mass of 3.7 $\times$ 10$^9$ \msun\ at a radius of 6.1 kpc.
We derived a total \hi\ mass of 1.3 $\times$ 10$^8$ \msun\ for ESO 324-G024 from integrating the \hi\ intensities and we find an average \hi\ mass surface density, $\Sigma_g$, of 8.8 M$_{\odot}$ pc$^{-2}$ in the disk and 1.7 M$_{\odot}$ pc$^{-2}$ in the tail.

\subsection{Tidal Force versus Ram Pressure}
 We calculate a tidal radius, $r_{\rm t}$ = 11 kpc for ESO 324-G024, which is nearly double the maximum \hi\ radius and thus, tidal forces do not appear to be the cause of the \hi\ tail in ESO 324-G024.
ESO 324-G024 has only one tail in its gas disk rather than two symmetric tails as would be expected if tidal forces were present.
ESO 324-G024 has a slightly positive peculiar velocity, which suggests that it is receding from our line of sight and implies that the galaxy may have passed through the northern radio lobe of Cen A.
The minimum IGM density required to create the \hi\ tail in ESO 324-G024 is $\rho_{\rm IGM}^{min} = 10^{-3}$ cm$^{-3}$ if the lobe IGM is at rest.
Assuming a lobe density of
10$^{-4}$ cm$^{-3}$ from \citet{osu13}, the required total velocity is $v_{\rm tot}$ = 330 \kms, which requires a minimal lobe speed of $v_{\rm lobe}$ = 197 \kms.
We estimate the minimum timescale, $\tau$ = 63 Myr, over which the ram pressure stripping has occurred. \\

In conclusion, we find that the most likely mechanism for creating the \hi\ tail in ESO 324-G024 is ram pressure stripping as the dIrr galaxy passed through the northern radio lobe of NGC 5128.  

\section{Acknowledgments}
This work is based on observations with the Australia Telescope
Compact Array (ATCA), which is operated by the CSIRO
Australia Telescope National Facility. This research has made use of the NASA/IPAC Extragalactic Database (NED) which is operated by the Jet Propulsion Laboratory, California Institute of Technology, under contract with the National Aeronautics and Space Administration.

\end{document}